\documentclass[fleqn,usenatbib,onecolumn]{mnras}

\usepackage{newtxtext,newtxmath}

\usepackage[T1]{fontenc}
\usepackage{ae,aecompl}

\usepackage{graphicx}	
\usepackage{amsmath}	
\usepackage{amssymb}	
\usepackage{bm}

\newcommand{\beq}{\begin{equation}}
\newcommand{\beqa}{\begin{eqnarray}}
\newcommand{\eeq}{\end{equation}}
\newcommand{\eeqa}{\end{eqnarray}}

\newcommand{\bfk}{\bm k}
\newcommand{\bftheta}{\bm \theta}
\newcommand{\bx}{{\bf x}}

\newcommand{\bk}{{\bf k}}
\newcommand{\bq}{{\bf q}}

\newcommand{\tSigma}{\tilde{\Sigma}}
\newcommand{\tdeltah}{\tilde{\delta}^{\rm 2D}_{\rm h}}

\newcommand{\pdeltahW}{{\delta}^{\rm 2D}_{{\rm h}W}}
\newcommand{\pdeltamW}{{\delta}^{\rm 2D}_{{\rm m}W}}
\newcommand{\deltah}{{\delta}_{\rm h}}
\newcommand{\deltam}{{\delta}^{\rm 2D}_{\rm m}}

\newcommand{\tdelta}{\widetilde{\delta}}
\newcommand{\tdeltaWX}{\widetilde{\delta}^{\rm 2D}_{{\rm X}W}}
\newcommand{\tdeltaWY}{\widetilde{\delta}^{\rm 2D}_{{\rm Y}W}}
\newcommand{\tW}{\widetilde{W}}
\newcommand{\bh}{\mathrm{h}}
\newcommand{\dr}{\mathrm{d}}

\newcommand{\hNh}{\hat{n}^{\rm 2D}_{\rm h W}}
\newcommand{\bNh}{\bar{n}^{\rm 2D}_{\rm h}}
\newcommand{\deltaml}{\delta_{\mathrm{m,lin}}}
\newcommand{\deltab}{\delta_{\rm b}}

\newcommand{\simgt}{\lower.5ex\hbox{$\; \buildrel > \over \sim \;$}}
\newcommand{\simlt}{\lower.5ex\hbox{$\; \buildrel < \over \sim \;$}}

\def\avrg#1{\left\langle #1 \right\rangle}

\title[Covariances for the projected correlations in the 
response approach]{Covariances for cosmic shear and galaxy--galaxy lensing in the response approach}

\author[R.~Takahashi et al.]{Ryuichi Takahashi$^{1}$, 
Takahiro Nishimichi$^{2}$, 
Masahiro Takada$^{2}$, 
Masato Shirasaki$^{3}$,
\newauthor
Kosei Shiroyama$^{1}$
\\
$^{1}$Faculty of Science and Technology, Hirosaki University, 3 Bunkyo-cho, Hirosaki, Aomori 036-8588, Japan\\
$^{2}$Kavli Institute for the Physics and Mathematics of the Universe (WPI), The University of Tokyo Institutes for Advanced Study (UTIAS),\\
The University of Tokyo, 5-1-5 Kashiwanoha, Kashiwa-shi, Chiba, 277-8583, Japan\\
$^{3}$National Astronomical Observatory of Japan, Mitaka, Tokyo 181-8588, Japan
}

\date{Accepted XXX. Received YYY; in original form ZZZ}

\pubyear{2017}

\begin{document}
\setlength{\mathindent}{0pt}
\label{firstpage}
\pagerange{\pageref{firstpage}--\pageref{lastpage}}
\maketitle

\begin{abstract}
In this study, we measure the response of matter and halo projected power spectra 
$P^{\rm 2D}_{\rm XY}(k)$ (X, Y are matter and/or halos), to a large-scale density contrast, $\delta_{\rm b}$, using separate universe simulations. 
We show that the fractional response functions, i.e., $\mathrm{d}\ln P^{\rm 2D}_{\rm XY}(k)/\mathrm{d}\delta_{\rm b}$, are identical to their respective three-dimensional power spectra within simulation measurement errors.
Then, using various $N$-body simulation combinations (small-box simulations with periodic boundary conditions and sub-volumes of large-box simulations) to construct {mock observations of projected fields}, we study how super-survey modes, in both parallel and perpendicular directions to the projection direction, affect the covariance matrix of $P^{\rm 2D}_{\rm XY}(k)$, known as super-sample covariance (SSC). 
Our results indicate that the SSC term provides dominant contributions to the covariances of matter-matter and matter-halo spectra at small scales but does not provide significant contributions in the halo-halo spectrum. 
We observe that the large-scale density contrast in each redshift shell causes most of the SSC effect, and we did not observe a SSC signature arising from large-scale tidal field within the levels of measurement accuracy.
We also develop a response approach to calibrate the SSC term for cosmic shear correlation function and galaxy--galaxy weak lensing, and validate the method by comparison with the light-cone ray-tracing simulations. 
Our method provides a reasonably accurate, albeit computationally inexpensive, way to calibrate the covariance matrix for clustering observables available from wide-area galaxy surveys. 
\end{abstract}




\section{Introduction} 
\label{sec:intro}

Weak gravitational lensing is a powerful cosmological probe to constrain the nature of dark matter and dark energy \citep[e.g.,][for a review]{Hoekstra:2008,Munshi:2008,Kilbinger:2015,Mandelbaum:2017}.
Measuring angular correlations between distant galaxy shapes enables the mapping of foreground matter distribution (i.e., {\em cosmic shear}).
Similarly, stacking shapes of background galaxies {around} foreground large-scale structure tracers, such as galaxies and clusters, provide reconstructions of the averaged matter distribution around the tracers
 (i.e., {\em galaxy--galaxy lensing} or {\em stacked cluster lensing}).
The weak lensing observables are sensitive to cosmological parameters such as a combination of the cosmological matter density ($\Omega_{\rm m}$) and the current amplitude of density fluctuations at $8 h^{-1}$Mpc scale ($\sigma_8$).
Several galaxy imaging surveys have put stringent constraints on $\Omega_{\rm m}$ and $\sigma_8$ from their cosmic shear measurements such as the results from 
the Canada-France-Hawaii Telescope Lensing Surveys\footnote{\url{http://www.cfhtlens.org}} \citep[CFHTLenS:][]{Heymans:2013,Kilbinger:2013}, the Dark Energy Survey\footnote{\url{http://www.darkenergysurvey.org}} \citep[DES:][]{DESCollaboration:2017,Troxel:2017}, and the Kilo-Degree 
Survey\footnote{\url{http://kids.strw.leidenuniv.nl}} \citep[KiDS:][]{Hildebrandt:2017}.
A combination of the Sloan Digital Sky Survey\footnote{\url{http://www.sdss.org}} (SDSS) spectroscopic galaxy catalogue with the galaxy--galaxy lensing measurement was used to obtain stringent cosmological constraints \citep[e.g.,][]{Mandelbaum:2013,Moreetal:2015}.
On-going and upcoming surveys intend to advance the weak lensing cosmology; the Subaru Hyper Suprime-Cam survey\footnote{\url{http://hsc.mtk.nao.ac.jp/ssp/}} \citep{HSCOverview:17}, the Large Synoptic Survey Telescope (LSST) project\footnote{\url{http://www.lsst.org}}, 
the ESA Euclid mission\footnote{\url{http://sci.esa.int/euclid/}} and the NASA WFIRST mission\footnote{\url{http://wfirst.gsfc.nasa.gov}}. 

To perform robust cosmological analysis with on-going and upcoming wide-area weak lensing surveys, we need accurate covariance matrix estimations for weak lensing observables, which describe statistical uncertainties in the measurements. 
The covariance matrix is divided into the following three contributions that vary in their statistical nature: {i)} the Gaussian covariance {based on the Gaussian field approximation},  {ii)} the {\it connected} non-Gaussian contribution from {the mode-coupling of density fluctuations inside the survey region (described by the trispectrum of sub-survey modes)} \citep{Scoccimarroetal:99,CoorayHu:01}, and {iii)} the super-sample covariance (SSC) contribution {from the correlation between sub-survey modes and super-survey modes with wavelengths larger than the survey region (described by the squeezed trispectrum)} {(\citealt{TakadaHu:13}, see also \citealt{HuKravtsov:03,Hamiltonetal:06}, for pioneer work)}.
The literature describes several methods to calibrate covariance matrices. 
The first uses a large number of mock catalogues constructed from numerical simulations, but these need to include the SSC contribution using a carefully designed simulation setup (e.g., simulations with box sizes larger than the survey volume) \citep[e.g.,][]{Satoetal:09,Satoetal:11,Takahashietal:09,HVvW:12,Blot:2015,Shirasakietal:17,Takahashietal:17}.
This method provides the most accurate calibration, but such a calibration has been done only for a single fiducial cosmological model. In addition, it is not generally straightforward to design simulated mock catalogues that can meet the following two competing requirements: 1) a wide-area light-cone volume that covers the large survey volume and 2) a {high} spatial resolution that accurately models the nonlinear structure formation. 
Therefore, this method is usually computationally expensive, even with currently available technological resources. 
If a data vector dimension increases, e.g., for combined probes of cosmic shear and galaxy--galaxy lensing, limitations on the number of mocks become severe \citep[e.g.,][]{Hartlap:2007,Taylor:2013}.
The second method estimates the covariance from the actual data itself (i.e., jackknife or bootstrap), where the survey region is divided into small sub-regions and the covariance can be  estimated from sub-region measurements \citep[see Section~2 in ][]{Norberg:2009}. 
However, the covariance matrix estimated in this method may suffer from noise owing to a limited number of realisations as well as not being able to estimate covariance at scales greater than the subregion size itself \citep[e.g.,][]{Miyatake:2015,Shirasakietal:17,Singhetal:16}, which prevents access to clustering signals at large scales. 
{\cite{Friedrich:2016} also demonstrated, for cosmic shear, that this method does not accurately estimate the sample variance but only the shape-noise variance.}
The third uses an analytical method or a hybrid method combining analytical and simulation-based methods; for example, the perturbation theory
\citep[e.g.,][]{MohammedSeljak:14,Mohammedetal:17}, {approximate methods for fast generating mock catalogues \citep[e.g.,][]{Chuang:2015,Lippich:2018}}, the halo model \citep[e.g.,][]{CoorayHu:01,TakadaBridle:07,TakadaJain:09,Satoetal:09,Kayoetal:13,TakadaHu:13,HikageOguri:16} or the response approach combined with simulations \citep{Lietal:14,Lietal:14b,2017arXiv171107467B,2018PASJ...70S..25M,Murataetal:17}. 
{\cite{Izard:2018} generated all-sky weak-lensing maps using an approximate $N$-body method \texttt{ICE-COLA}.}
{\cite{Prat:2017} and \cite{Troxel:2017} prepared log-normal weak-lensing mocks for data analysis in DES.} 
\cite{KrauseEifler:16} developed an analytical covariance formula based on the halo model, which includes the SSC contribution, adopted for cosmological analyses of the first-year DES data \citep{DESCollaboration:2017} and the KiDS data \citep{Hildebrandt:2017}.

The purpose of this study is to develop a SSC calibration method for projected matter-matter and matter-halo power spectra that are relevant for cosmic shear and galaxy--galaxy weak lensing, respectively. To develop this method, we first calibrate the response functions of the projected power spectra to the large-scale density contrast using separate universe simulations \citep{Sirko:2005,Lietal:14,Lietal:14b,Baldauf:2011,Wagner:2015,Baldaufetal:16,BarreiraSchmidt:17,Chiang:17,Schmidt:2018}, where the effects of large-scale density contrast are absorbed by changes in background cosmological parameters used in $N$-body simulations. 
We study whether the response functions of the projected power spectra differ from those of their respective three-dimensional power spectra. 
In observing these spectra, we pay particular attention to the manner in which super-survey modes, both in parallel and perpendicular directions to the projection direction, affect the projected power spectra using mock catalogues of projected fields that are
 constructed from a combination of different $N$-body simulations. 
Subsequently, we develop a method to calibrate SSC contributions for the cosmic shear correlation function and the halo-convergence cross-correlation function using response functions calibrated from separate universe simulations. We validate this method 
by comparing it with the results from light-cone ray-tracing simulations. 
Then, we discuss the approach to calibrate the covariance matrix by combining 
the actual data, the response approach, and {numerical} simulations. 

Throughout this study, we adopt the flat-geometry $\Lambda$CDM (Lambda cold dark matter) model that is consistent with the Planck 2015 results \citep[][hereafter, {\it Planck}]{Planck:2015}. 
The cosmological parameters are as follows: the matter density, $\Omega_{\rm m}=1-\Omega_{\Lambda}=0.3156$, the baryon density, $\Omega_{\rm b}=0.0492$, the Hubble parameter, $h=H_0/(100 {\rm km} \, {\rm s}^{-1} \, {\rm Mpc}^{-1}) = 0.6727$, the present amplitude of density contrast at $8 \, h^{-1} \, {\rm Mpc}$, $\sigma_8=0.831$, and the spectral index, $n_{\rm s}=0.9645$.  
{We use units of $c=G=1$.}

\section{Power spectrum covariance of projected field in a finite survey area}

\subsection{Projected halo number density field}
\label{sec:projected_field}

In this subsection, we define the statistical quantities of the halos. 
We assume that halos in a survey region are identifiable via observables such as optical richness and X-ray observables; we also assume that mass and redshift of each halo are available. 
However, the following discussion can be extended to a general case in which only their proxies are available.

The number density field for halos, in the mass range of $[M,M+\dr M]$, at {a two-dimensional position $\bx$ perpendicular to the line-of-sight} and redshift $z$, is defined as 
\begin{equation}
 {\frac{\dr n_{\rm h}}{\dr M} (\bx, z; M) = \frac{\dr n}{\dr M} (M,z) 
 \left[1+\deltah(\bx, z; M)\right],}
\end{equation}
where $\mathrm{d}n/\mathrm{d}M$ is the (ensemble-averaged) number density of halos with mass $[M,M+\dr M]$, ${\deltah(\bx, z; M)}$ is the three-dimensional number density fluctuation field for halos at a redshift of $z$ with mass $M$.
{The comoving radial distance, $\chi$,} is given as a function of redshift via the distance-redshift relation, $\chi=\chi(z)$, for the underlying true cosmological model.
Note that we can infer the projected position vector via $\bx=\chi(z) \btheta$ for a flat geometry universe, where $\bftheta$ is the angular position vector.

By denoting a radial weight function, $f_\bh(\chi)$, and a survey window function, $W(\bx)$, we can define the projected number density field for halos, integrated over a range of redshift and halo masses, in terms of the three-dimensional density field, {$[\dr n_{\rm h}/\dr M](\bx, z; M)$}, as follows:
\begin{align}
n^{\rm 2D}_{{\rm h}W}(\bx) &= {\int\!\dr\chi \, f_\bh(\chi) \, W(\bx) \int\!\dr M \, \frac{\dr n_\bh}{\dr M}(\bx, \chi; M)} ,\notag \\
 &= {\int\!\dr\chi \, f_\bh(\chi) \, W(\bx) \int\!\dr M \frac{\dr n}{\dr M}(M,\chi) \left[1+\deltah(\bx, \chi; M)\right]}.
  \label{eq:Nh}
\end{align}
Here, $f_\bh(\chi)$ is the radial weight or selection function, given as a function of redshift (or $\chi$), e.g., $f_\bh(\chi)\ne 0$, if $\chi$ is inside the redshift range of a survey, otherwise $f_\bh(\chi)=0$. 
Similarly, the survey-window function, $W(\bx)$, is defined to satisfy the condition $W(\bx)=1$ if the position vector $\bx$ is inside a survey region, otherwise $W(\bx)=0$. 
We use the subscript ``$W$'' to denote the field measured in a survey region, and use the superscript ``2D'' to denote the projected quantities.
Note that, if $f_{\rm h}(\chi)$ is dimensionless, $n^{\rm 2D}_{{\rm h}W}$ has a dimension of $({\rm length})^{-2}$.
We use a distant observer approximation so that the survey boundary does not depend on the redshift.
This is a good approximation for galaxy--galaxy lensing, where the lensing galaxies are taken from a narrow redshift width. 
This approximation is also valid for cosmic shear if radial integration is replaced by a discrete summation on thin lens shells (i.e., $\int \! \dr \chi \simeq \sum_i \Delta \chi$, where $\Delta \chi$ is the thickness of the lens shell) for each shell that we can use the approximation, $\bx=\chi_i\bftheta$, where $\chi_i$ is the mean distance to the $i$-th lens shell.
The projected survey area, $S_W$, is given by
\begin{equation}
S_W=\int\!\!\dr^2 \bx \, W(\bx).
\end{equation}

An estimator of the {\em mean} projected halo number density, in a survey volume, is defined as:
\begin{align}
 \hNh&=\frac{1}{S_W}\int\!\dr^2\bx \, n^{\rm 2D}_{{\rm h}W}(\bx), \nonumber\\
 &= \int\!\dr\chi \, f_\bh(\chi)\int\!\dr M \, \frac{\dr n}{\dr M}(M,\chi) \left[ 1+\frac{1}{S_W}\int\!\dr^2\bx \, W(\bx) \, \deltah(\bx,\chi;M) \right], \nonumber\\
 &\simeq \int\!\dr\chi \, f_\bh(\chi)\int\!\dr M \, \frac{\dr n}{\dr M}(M,\chi) \left[ 1+b(M,\chi) \, \frac{1}{S_W}\int\!\dr^2\bx \, W(\bx) \, \deltaml(\bx,\chi) \right],\nonumber\\
&\simeq \bNh\left[1+\bar{b}_\bh \, \deltab\right],
\label{eq:hNh}
\end{align}
where 
\begin{align}
\bNh &\equiv \int\!\dr\chi \, f_\bh(\chi)\int\!\dr M \, \frac{\dr n}{\dr M}(M,\chi), \nonumber\\
\bar{b}_\bh &\equiv \frac{1}{\int\!\dr M^\prime \, \frac{\dr n}{\dr M}(M^\prime,\chi)}\int\!\dr M \, \frac{\dr n}{\dr M}(M,\chi) \, b(M), \nonumber\\
\deltab &\equiv \frac{1}{\int\!\dr\chi^\prime \, f_\bh(\chi^\prime) \int\!\dr^2\bx^\prime \, W(\bx^\prime)} \int\!\dr\chi \, f_\bh(\chi) \int\!\dr^2\bx \, W(\bx) \, \deltaml(\bx,\chi).
\label{halo_propaties}
\end{align}
Here, we use a halo density fluctuation field at large scales, which is given as
$\delta_\bh\simeq b(M)\deltaml$, where $b(M)$ is the linear halo bias parameter with mass $M$ and $\deltaml$ is the linear mass density fluctuation field. 
We also assume that both the halo mass function and halo bias are not rapidly varying functions of redshift; therefore, we ignore their redshift dependences in the radial integration. 
In the above equations, $\bNh$ is the ensemble average of projected number density,
$\bar{b}_\bh$ is the mean halo bias in the sample, and $\deltab$ is the average density contrast within the survey volume. 
Equation~(\ref{eq:hNh}) {indicates} that the number of halos in a survey region is generally modulated by the large-scale density contrast, $\deltab$, and up-weighted by the halo bias, ${\bar{b}_\bh}$, on the basis of an individual survey region \citep{HuKravtsov:03,TakadaBridle:07}. 

The projected halo density fluctuation field is defined {from equations (\ref{eq:Nh}) and (\ref{eq:hNh})} as 
\begin{equation}
\pdeltahW(\bx)\equiv \frac{1}{\hNh} \left[n^{\rm 2D}_{{\rm h}W}(\bx) - \hNh \right] {=\frac{1}{\hNh}\int\!\dr\chi \, f_\bh(\chi) \, W(\bx) \int\!\dr M \, \frac{\dr n}{\dr M}(M,\chi) \, \deltah(\bx,\chi; M).}
\label{eq:2Ddeltah}
\end{equation}
Similarly, we can define the projected mass density fluctuation field as 
\begin{equation}
\pdeltamW(\bx)\equiv 
\int\!\dr\chi \, f_{\rm m}(\chi) \, W(\bx) \, \delta_{\rm m}(\bx,\chi),
\label{eq:2Ddeltam}
\end{equation}
where $f_{\rm m}(\chi)$ is the radial weight function that depends only on $\chi$ ($f_{\rm m}$ may have a dimension, but here we keep it general). 
If we take an appropriate form of $f_{\rm m}(\chi)$, we can express the weak lensing or cosmic shear field using the form above. 
Unlike in equation~(\ref{eq:2Ddeltah}), we do not normalise the projected matter field by the 
local mass density contrast, $\deltab$, because weak lensing measured from the lensing distortion effects on galaxy shapes arises from gravitational potential fields in large-scale structures, related to matter density fluctuation fields with respect to 
the {\em global background} mean mass density \citep[see][for details]{TakadaHu:13,Lietal:14}.

\subsection{An estimator of projected power spectrum}
\label{sec:estimator}

From {equations}~(\ref{eq:2Ddeltah}) and (\ref{eq:2Ddeltam}), we can define a general form to express the projected field of matter or halos as
\begin{equation}
\delta^{\rm 2D}_{{\rm X}W} (\bx)\equiv \int\!\mathrm{d}\chi \, F_{\rm X}(\chi) \, W(\bx) \, \delta_{\rm X}(\bx,\chi),
\end{equation}
where the subscript ``X'' denotes either matter or halo (${\rm X}={\rm m}$ or h), and $F_{\rm X}(\chi)$ is a radial function, e.g., $F_{\rm X}(\chi)=f_{\rm m}(\chi)$ or $F_{\rm X}(\chi)=[f_{\rm h}(\chi)/\hNh] \int \!\dr M\, [\mathrm{d}n/\mathrm{d}M](M,\chi)$ for matter or halos, respectively.
The Fourier transform of the projected field is represented as 
\begin{equation}
\tdeltaWX(\bk_\perp)= {\int\!\mathrm{d}\chi \, F_{\rm X}(\chi)\int\!\frac{\mathrm{d}^2\bq_\perp}{(2\pi)^2} \, \tW(\bq_\perp)
\int \!\frac{\mathrm{d}q_\parallel}{2\pi} \, \tdelta_{\rm X}(\bk_\perp \!  -\bq_\perp,q_\parallel;\chi) \, e^{iq_\parallel\chi},}
\end{equation}
where quantities with tilde symbols, ``$\widetilde{\hspace{2mm}}$'', denote the Fourier transforms, $\bk_\perp$ and $\bq_\perp$ are two-dimensional wavevectors perpendicular to the line-of-sight direction, and $k_\parallel$ {and $q_\parallel$ are the parallel components}. 
We often ignore the subscript, ``${}_\perp$'', to denote the perpendicular vector for notational simplicity.

Let us then define a power spectrum estimator of the projected field as follows:
\begin{equation}
\widehat{P}_{{\rm XY}W}^{\rm 2D}(k) \equiv \frac{1}{S_W}\int_{|\bk^\prime|\in k}\frac{\mathrm{d}^2\bk^\prime}{S_{k}} \, {\rm Re} \left[ \tdeltaWX(\bk^\prime) \, \tdeltaWY(-\bk^\prime) \right],
\label{eq:pk_est}
\end{equation}
where we integrate over a circular annulus of width $\Delta k$, around the radius $k$ in two-dimensional $k$-space, and $S_{k}\equiv \int_{|\bk^\prime|\in k}\!\mathrm{d}^2\bk'\simeq 
2\pi k\Delta k$ for $k\gg 1/\sqrt{S_W}$.
The ensemble average of the power spectrum estimator (equation~(\ref{eq:pk_est})) is computed as 
\begin{equation}
\avrg{\widehat{P}_{{\rm XY}W}^{\rm 2D}(k)}=\frac{1}{S_W}\int_{|\bk^\prime|\in k}\!\frac{\mathrm{d}^2\bk^\prime}{S_{k}}\int\!\frac{\mathrm{d}^2\bq}{(2\pi)^2}
\left| \, \tW(\bq) \, \right|^2 P^{\rm 2D}_{\rm XY}(|\bk^\prime-\bq|).
\end{equation}
Here, $P^{\rm 2D}_{\rm XY}(k)$ is the {power spectrum of projected field} that is expressed in terms of the underlying three-dimensional power spectrum using Limber's approximation \citep{Limber:54}:
\begin{equation}
P^{\rm 2D}_{\rm XY}(k)\simeq \int\!\mathrm{d}\chi \, F_{\rm X}(\chi) \, F_{\rm Y}(\chi) \, P_{\rm XY}(k;\chi), 
\label{eq:pk2d_def}
\end{equation}
where $P_{\rm XY}(k;\chi)$ is the three-dimensional power spectrum, defined as
\begin{equation}
\avrg{\tdelta_{\rm X}(\bk,k_\parallel;\chi) \, \tdelta_{\rm Y}(\bk^\prime,k_\parallel^{\prime};\chi)} \equiv (2\pi)^3\delta^2_{\rm D}(\bk+\bk') \, \delta_{\rm D}(k_\parallel+k_\parallel^{\prime}) \, P_{\rm XY}\!\left(\sqrt{k^2+k_{\parallel}^2};\chi\right),
\end{equation}
where $\delta_{\rm D}(k)$ is the Dirac delta function. In equation~(\ref{eq:pk2d_def}), we assume that the 
{correlation function of projected field} 
originates predominantly from correlations in the underlying three-dimensional fields at {\it equal} times of $\chi=\chi^\prime$ such that Fourier modes are perpendicular to the line-of-sight direction and that effects from the radial Fourier mode are negligible. 
This is a good approximation for weak lensing statistics \citep{ValeWhite:03}, including galaxy--galaxy lensing or stacked cluster lensing. 
{Formulation in this study are extendable to include the effects from the radial mode using \cite{KitchingHeavens:17}}.

In this study, we are particularly interested in the effects {of} global survey geometry on the covariance of the projected power spectrum and not in the effects of {small-scale features in the mask}. 
If we focus on the wavenumber modes, which satisfy $k \gg S_W^{-1/2}$, we observe that the power spectrum estimator (equation~(\ref{eq:pk_est})) is unbiased:
\begin{align}
\avrg{\widehat{P}_{{\rm XY}W}^{\rm 2D}(k)} &\simeq \frac{1}{S_W}\int_{|\bk^\prime|\in k}\!\frac{\mathrm{d}^2\bk^\prime}{S_k}P^{\rm 2D}_{\rm XY}(k^\prime)
\int\!\frac{\mathrm{d}^2\bq}{(2\pi)^2}\left| \, \tW(\bq) \, \right|^2, \nonumber\\
&\simeq P^{\rm 2D}_{\rm XY}(k)\frac{1}{S_W}
\int\!\frac{\mathrm{d}^2\bq}{(2\pi)^2}\left| \, \tW(\bq) \, \right|^2, \nonumber\\
&= P^{\rm 2D}_{\rm XY}(k).
\end{align}
Here, we assume that $P^{\rm 2D}_{\rm XY}(|\bk-\bq|)\simeq P^{\rm 2D}_{\rm XY}(k)$ over an integration range of $\bq$ {because $\tW(\bq)$ has a peak at $q \sim S_W^{-1/2}$ which is much smaller than $k$.}
{We} also assume that $P^{\rm 2D}_{\rm XY}(k)$ is not a rapidly varying function within the $k$-bin.
In the above equation, we have used $S_W=\int|\tW(\bq)|^2\mathrm{d}^2\bq/(2\pi)^2$ for the window function.

\subsection{Covariance of the projected power spectrum}
\label{sec:cov_derive}

The covariance matrix of the projected power spectrum can be formally expressed as
\begin{align}
{\rm Cov}[P^{\rm 2D}_{\rm XY}(k),P^{\rm 2D}_{\rm XY}(k')]
&\equiv \avrg{\widehat{P}^{\rm 2D}_{\rm XY}(k)\widehat{P}^{\rm 2D}_{\rm XY}(k')}
-\avrg{\widehat{P}^{\rm 2D}_{\rm XY}(k)}\avrg{\widehat{P}^{\rm 2D}_{\rm XY}(k')}\nonumber\\
&= {\rm  Cov}^{\rm G}(k,k')+{\rm Cov}^{\rm cNG}(k,k')+{\rm Cov}^{\rm SSC}(k,k').
\label{eq:cov_def}
\end{align}
The covariance matrix is broken down into three parts: the Gaussian (``G'') covariance, the connected non-Gaussian (``cNG'') covariance and the super-sample covariance (``SSC''). 
The {G} and cNG terms scale with the survey area as $S_W^{-1}$, but the SSC term does not follow this simple scaling. 
{The G and cNG terms will be discussed in later sections \ref{sec:covariance}, \ref{sec:cosmic_shear}, \ref{sec:gg_lens}, and \ref{sec:conclusion}.}

Next, we derive the expression for the SSC term using the response function of the three-dimensional (3D) power spectrum to the large-scale density contrast along the line-of-sight direction.
We express the radial integration in the projected power spectrum  (equation~(\ref{eq:pk2d_def})), using a discrete summation of the $N_{\rm sub}$ radial shells, as
\begin{equation}
\widehat{P}^{\rm 2D}_{\rm XY}(k)\simeq\sum_{i=1}^{N_{\rm sub}}\Delta\!\chi \, F_{\rm X}(\chi_i)F_{\rm Y}(\chi_i)
\widehat{P}_{\rm XY}\!\left(k;\chi_i\right),
\label{eq:P2d_discrete}
\end{equation}
where $\chi_i$ is the {comoving} distance to the $i$-th redshift shell (e.g., the distance to the central redshift of the shell), $\Delta\!\chi$ is the width of the shell, and we assume that all shells have the same {comoving} width.  
{A discrete summation is a good approximation because both the power spectrum and the radial weight are not a rapidly varying function of redshift within the shell.}
In the presence of super-survey modes in each shell, the power spectrum estimator modulates to:
\begin{align}
\widehat{P}_{\rm XY}^{\rm 2D}(k) &\simeq \sum_{i=1}^{N_{\rm sub}}\Delta\!\chi \, F_{\rm X}(\chi_i)F_{\rm Y}(\chi_i)\widehat{P}_{\rm XY}\!\left(k;\delta_{{\rm b}i},\chi_i\right)\nonumber\\
&\simeq \sum_{i=1}^{N_{\rm sub}}\Delta\!\chi \, F_{\rm X}(\chi_i)F_{\rm Y}(\chi_i)
\left[\widehat{P}_{\rm XY}\!\left(k;\delta_{{\rm b}i}=0,\chi_i\right)+\left.\frac{\partial P_{\rm XY}(k;\delta_{{\rm b}i},\chi_i)}{\partial \deltab}\right|_{\delta_{{\rm b}i}=0}
\delta_{{\rm b}i}\right],
\end{align}  
where $\partial P_{\rm XY}/\partial \deltab$ is the power spectrum response to large-scale density modes \citep{TakadaHu:13,Lietal:14}, and $\delta_{{\rm b}i}$ is the super-survey mode of the $i$-th shell. 
For each particular survey realisation, each $\delta_{{\rm b}i}$ has a particular value.   
We calculate the SSC contribution to the projected power spectrum covariance with the following equation:
\begin{align}
{\rm Cov}^{\rm SSC}\left[P^{\rm 2D}_{\rm XY}(k),P^{\rm 2D}_{\rm XY}(k')\right]&=
\sum_{i,j}(\Delta\!\chi)^2 \, F_{\rm X}(\chi_i) \, F_{\rm X}(\chi_j) \, F_{\rm Y}(\chi_i) \, F_{\rm Y}(\chi_j) \frac{\partial P_{\rm XY}(k;\chi_i)}{\partial \deltab} \frac{\partial P_{\rm XY}(k';\chi_j)}{\partial \deltab} \avrg{\delta_{{\rm b}i}\delta_{{\rm b}j}}\nonumber\\
&\simeq 
\sum_{i=1}^{N_{\rm sub}}(\Delta\!\chi)^2 \, F_{\rm X}^2(\chi_i) \, F_{\rm Y}^2(\chi_i) \frac{\partial P_{\rm XY}(k;\chi_i)}{\partial \deltab} \frac{\partial P_{\rm XY}(k';\chi_i)}{\partial \deltab} \sigma_{\rm b}^2(z_i),
\label{eq:ssc_derivation}
\end{align}
where we assume $\avrg{\delta_{{\rm b}i}}=0$ and
$\avrg{\delta_{{\rm b}i}\delta_{{\rm b}j}}=\sigma_{\rm b}^2(z_i) \, \delta^{\rm K}_{ij}$. For the latter condition, we assume that the super-survey modes between different shells are uncorrelated with each other. In other words, we ignore radial large-scale modes that cause correlations between the different $\delta_{{\rm b}i}$. We will verify this assumption in Section \ref{sec:covariance}.

When we appropriately consider redshift evolution within the SSC computation, we use Limber's equation to derive a line-of-sight integration from the discrete summation formula above:
\begin{equation}
\sum_{i=1}^{N_{\rm sub}}(\Delta\!\chi)^2 \, F_{\rm X}^2(\chi_i)F_{\rm Y}^2(\chi_i) \frac{\partial P_{\rm XY}(k;\chi_i)}{\partial \deltab} \frac{\partial P_{\rm XY}(k';\chi_i)}{\partial \deltab} \sigma_{\rm b}^2(z_i)
\simeq \int\!\!\mathrm{d}\chi \, F_{\rm X}^2(\chi)F_{\rm Y}^2(\chi) \frac{\partial P_{\rm XY}(k;\chi)}{\partial \deltab} \frac{\partial P_{\rm XY}(k';\chi)}{\partial \deltab} \sigma_{\rm b}^2(\chi) \Delta\!\chi \, .
\end{equation}
Following \citet[][]{TakadaSpergel:14} (the discussion near equation~(33) in their study), we find that: 
\begin{align}
\sigma_{\rm b}^2(\chi)\Delta\!\chi &= \avrg{\delta_b^2} \Delta \chi \nonumber \\
&=\frac{\Delta\!\chi}{(S_W\Delta\!\chi)^2}\int\!\frac{\mathrm{d}^2\bk_\perp} {(2\pi)^2}\int\!\frac{\mathrm{d}k_\parallel}{2\pi} P_{\rm m,lin}\!\left(k=\sqrt{k_\parallel^2+k_\perp^2};\chi\right)\left| \, \tW_\perp(\bk_\perp) \, \right|^2\left| \, \tW_\parallel(k_\parallel) \, \right|^2\nonumber\\
&\simeq \frac{1}{(S_W)^2\Delta\!\chi}
\int\!\frac{\mathrm{d}^2\bk_\perp}{(2\pi)^2} P_{\rm m,lin}\!\left(k_\perp;\chi\right)\left| \, \tW_\perp(\bk_\perp) \, \right|^2 \left[
\int\!\frac{\mathrm{d}k_\parallel}{2\pi} \left| \, \tW_\parallel(k_\parallel) \, \right|^2\right] \nonumber\\
&\simeq \frac{1}{(S_W)^2} \int\!\frac{\mathrm{d}^2\bk_\perp}{(2\pi)^2}~ P_{\rm m,lin}\!\left(k_\perp;\chi\right)\left| \, \tW_\perp(\bk_\perp) \, \right|^2,
\label{eq:sigmab_W}
\end{align}
from equation~(\ref{halo_propaties}).
Here, $P_{\rm m,lin}$ is the three-dimensional power spectrum of linear matter density, 
$W_\perp(\bx_\perp)$ and $W_\parallel(\chi)$ are the survey windows for the shell in perpendicular and parallel directions, respectively,
and we use the approximation: $\Delta\!\chi \simeq \int\!\mathrm{d}\chi W_\parallel(\chi) = \int\!\mathrm{d}\chi W_\parallel^2(\chi)=\int\!(\mathrm{d}k_\parallel/2\pi)|\tW_\parallel(k_\parallel)|^2$.
%
Therefore, the SSC term is expressed as 
\begin{equation}
{\rm Cov}^{\rm SSC}\left[P^{\rm 2D}_{\rm XY}(k),P^{\rm 2D}_{\rm XY}(k^\prime)\right]
= \int\!\mathrm{d}\chi \, F_{\rm X}^2(\chi)F_{\rm Y}^2(\chi)
\frac{\partial P_{\rm XY}(k;\chi)}{\partial \deltab}
\frac{\partial P_{\rm XY}(k^\prime;\chi)}{\partial \deltab}
\frac{1}{(S_W)^2}\int\!\frac{\mathrm{d}^2\bq}{(2\pi)^2} \, P_{\rm m,lin} (q;\chi) \left| \, \tW_\perp(\bq) \, \right|^2.
\end{equation}
Hence, we are able to recover the standard SSC expression for the projected power spectrum \citep{Satoetal:09,TakadaHu:13,TakadaSpergel:14,Schaanetal:14}. 
The SSC term depends solely on the survey window originating from $\sigma_{\rm b}^2$.
For a sufficiently wide survey area we can use the linear matter power spectrum or Gaussian simulations to accurately compute $\sigma_{\rm b}^2$ for a given survey window.
We suggest that the SSC formulae are good approximations as far as the effects of the radial Fourier modes on the projected power spectrum are negligible. 
We will perform various simulations to validate the SSC formulae.

\section{Testing SSC of projected power spectrum with $N$-body simulations}
\label{sec:n-body}

In this section, we compare the covariance matrices from our analytical formula with those from {an ensemble of} $N$-body simulations. To do this, first, we numerically evaluate the ${P(k)}$ response to large-scale density contrasts using separate universe (SU) simulations (Section~\ref{sec:response}), and subsequently study SSC of ${P^{\rm 2D}(k)}$ by using a sufficient number of $N$-body realisations (Section~\ref{sec:covariance}).    

\subsection{Power spectrum response}
\label{sec:response}

\begin{table*}
	\centering
	\begin{tabular}{ccccccc} 
		\hline
	${\delta_{{\rm b}0}}$ & $L_W \, (h_{ W}^{-1} {\rm Mpc})$ & $h_{ W}$ & $\Omega_{{\rm m} W}$ & $\Omega_{\Lambda { W}}$ & $\Omega_{{\rm K} W}$ & $r_{\rm soft} \, (h_{W}^{-1}{\rm kpc})$ \\
		\hline
		$0$ & 250 & $0.6727$ & $0.3156$ & $0.6844$ & $0$ & $24.4$ \\
     $+0.01$ & $249.16$ & $0.67045$ & $0.3177$ & $0.6890$ & $-0.0067$ & $24.3$ \\
	 $-0.01$ & $250.83$ & $0.67494$ & $0.3135$ & $0.6799$ & $0.0066$ & $24.5$ \\
		\hline
	\end{tabular}
		\caption{The $N$-body simulation parameters listed for the fiducial cosmological model (second row) and SU simulations (third and fourth rows). The first column denotes the local density contrast at $z=0$, ${\delta_{{\rm b}0}}$, in the SU simulations (${\delta_{{\rm b}0}}=0$ for the fiducial model).				
		The second to the seventh columns list the simulation box size ($L_{\rm W}$), the Hubble parameter ($h_{\rm W}$), the matter density parameter ($\Omega_{\rm m W}$), the cosmological constant ($\Omega_{\Lambda {\rm W}}$), the curvature parameter ($\Omega_{\rm K W}$), and the softening length ($r_{\rm soft}$), respectively.
        {The number of particles is $512^3$ and the number of realisations is $100$ in each case.}
		}
	\label{table1}
\end{table*}

{The SU simulations provide the most accurate $P(k)$ responses even in the non-linear regime, compared to analytical methods. For instance, the perturbation theory is reliable only in the linear regime \citep{Baldaufetal:16} and the halo model also works in the non-linear regime but less accurate \citep{TakadaHu:13}.}
In {this SU} method, we absorb the large-scale density contrast, $\delta_{\rm b}$, into changes in cosmological parameters and subsequently study the effects on large-scale structures using $N$-body simulations under the changed cosmological background. 
{We give a brief summary of the SU method in Appendix A.}

We employ the {\it growth-dilation method} in \cite{Lietal:14} to implement the SU simulations. 
Hereafter, quantities with the subscript {``W''} denote quantities in the SU simulation whose value is different from that in the global background\footnote{{So far the script ``W'' means the window quantities but hereafter it means the SU quantities.}}.   
{Relations between quantities with and without ``W'' are discussed in Appendix A.}
We perform $100$ SU simulations for each super-survey mode, ${ \delta_{{\rm b}0} \equiv } \delta_{\rm b}(z=0)  =\pm 0.01$ (i.e., 100 paired simulations), where $\delta_{\rm b}$ evolves according to the spherical collapse dynamics.
The simulation and cosmological parameters are summarised in Table \ref{table1}.
We used a fixed number of $N$-body particles, $512^3$, but we changed the box size for each SU simulation (${\delta_{{\rm b}0}} =\pm 0.01$) so that the corresponding comoving lengths in the global background were identical: $L_W~(h_W^{-1}{\rm Mpc})=250~h^{-1}{\rm Mpc}=L~(h^{-1}{\rm Mpc})$. 
We adopted an identical initial seed for each paired SU (${\delta_{{\rm b}0}}=\pm0.01$) simulation to reduce contamination from sample variance. 
We employed the second-order Lagrangian perturbation theory \cite[2LPT;][]{Crocce:2006,Nishimichi:2009} to compute initial displacements at a redshift of $z=49$. We obtain the evolution of the scale factor, $a_W(t)$, in the SU comoving frame as a function of the global scale factor by solving the spherical collapse dynamics. Here, we do not linearise the equation so that it can still compute large $\delta_{\rm b}$ values in the $\Lambda$CDM global cosmological model.
The first- and second-order growth factors for the density perturbations are consistently computed in the SU background and subsequently are used to compute particle displacements.
Outputs of simulation data are dumped at two redshifts, $z_W$, for each SU simulation corresponding to $z=0.20$ and $0.55$ in the global background.
We followed the gravitational evolution of particles using \texttt{Gadget2}
\citep{Springel:2001,Springel:2005}, which we slightly modified to take into account the differences between local and global densities. The modified code reads a data table of the Hubble expansion rate, $H_W(t)$, in the SU background, and uses it {for the force calculation}. 
{For the \texttt{Gadget2} simulation parameters such as the time step and the force computation, we use the same parameters in our previous simulation \citep{Takahashi2012}, in which we checked that the matter power spectrum converges within $2 \, (6) \%$ for the wavenumber less than the particle Nyquist wavenumber ($\times 3$) corresponding to $k < 6 \, (20) \, h {\rm Mpc}^{-1}$ in the current setting.}

We also identify halos in an $N$-body simulation output using the public code \texttt{Rockstar} \citep{Behroozi:2013}, which we also slightly modified to make the resultant halo catalogues consistent among different frames. 
{The code first identifies groups of particles using the Friends-of-Friends(FOF) algorithm in six phase-space dimensions and then defines halos and subhalos in each group as spherical over-density regions.
The FOF linking length and density thresholds} are set to be the same when translated into the global frame. 
Throughout this study, we adopt $M_{\rm 200m}$ for the definition of halo mass {(i.e., the mean mass density of halo is $200$ times overdense compared with the global background matter density)}. 
In the following, we consider two halo samples, which were selected by their sharp mass thresholds, $M_{200 {\rm m}}>10^{12}$ or $>10^{13} h^{-1} \, {\rm M}_\odot$, respectively.

For each SU simulation realisation, we project the particle and halo distributions into two-dimensional $xy$-, $yz$- or $zx$-plane.
Then, we compute the projected density fluctuation fields of matter or halo as
\begin{align}
 \delta_{{\rm m}W}^{\rm 2D}(\bx_W) &= \frac{\rho^{\rm 2D}_{{\rm m}W}(\bx_W)}{\hat{\rho}^{\rm 2D}_{{\rm m}W}}-1 , \nonumber\\
 \delta^{\rm 2D}_{{\rm h} W}(\bx_W) &= \frac{n^{\rm 2D}_{{\rm h}W}(\bx_W)}{\hat{n}^{\rm 2D}_{{\rm h} W}}-1 ,
\label{eq:2Dfield_def}
\end{align}
where $\bx_W$ is the two-dimensional vector (in the $xy$-, $yz$- or $zx$-plane) {and the mean densities ($\hat{\rho}^{\rm 2D}_{{\rm m}W}$ and $\hat{n}^{\rm 2D}_{{\rm h} W}$) are measured in the SU simulation box.}
Then, we compute the Fourier components, $\tilde{\delta}^{\rm 2D}_{{\rm m}W}(\bfk_W)$ and $\tilde{\delta}^{\rm 2D}_{{\rm h}W}(\bfk_W)$, by performing a two-dimensional (2D) Fourier transform\footnote{We used the public code FFTW3 (Fast Fourier Transform in the West) at http://www.fftw.org/.} with $16384^2$ grid {cells, resulting in a maximum wavenumber $k_{\rm max} = (\pi/L) \times 16384 \simeq 206 \, h{\rm Mpc}^{-1}$}.
The power spectrum estimator is defined as 
\beq
  \hat{P}_{{\rm XY}W}^{\rm 2D}(k) = \frac{1}{N_k} \sum_{\bfk} {\rm Re} \left[ \tilde{\delta}_{{\rm X}W}^{{\rm 2D}*}(\bfk) \,  \tilde{\delta}_{{\rm Y}W}^{\rm 2D}(\bfk) \right],
\label{pk}
\eeq 
which gives the matter ${P^{\rm 2D}(k)}$ if the subscripts are ${\rm X}={\rm Y}={\rm m}$, the halo ${P^{\rm 2D}(k)}$ if ${\rm X}={\rm Y}={\rm h}$, and the halo-matter cross ${P^{\rm 2D}(k)}$ if ${\rm X}={\rm h}$ and ${\rm Y}={\rm m}$. 
The summation is done over the circular annulus $(k-\Delta k/2, k+\Delta k/2)$ with a bin-width of $\Delta k$ in 2D Fourier space, and $N_k$ is the number of modes in the $k$-bin, $N_{k} \simeq 2 \pi k \Delta k/(2\pi/L)^2\simeq k\Delta k S_W/(2\pi)$ for $k\gg 1/L$ {(here, $S_W=L^2$)}.
For the halo auto-power spectrum, we subtract the shot noise from the measured $\hat{P}_{{\rm hh}W}^{\rm 2D}(k)$ in equation~(\ref{pk}).
We also calculate the 3D power spectrum estimator, $\hat{P}_{{\rm XY}W}(k)$, following a similar procedure as described above.

We briefly mention a procedure to evaluate the 3D $P(k)$ response function following the methods described in \cite{Lietal:14}.
We use the 3D power spectrum measured in each SU simulation to infer the power spectrum in the global background via the relation
\beq 
 P_{\rm XY}(k) \, k^3 = \left( 1+\delta_{\rm b} \right)^n P_{{\rm XY} W}(k_{W}) \, k_{W}^3,
\label{pkresp_relation}
\eeq
where $k_{ W}=(1 - \delta_{\rm b}/3) \, k$ is a wavenumber measured in the SU simulation, 
and $n=2,1,$ and $0$ when ${\rm XY} = {\rm mm}, {\rm hm}$ and ${\rm hh}$, respectively. 
The extra factor $(1+\delta_{\rm b})^n$ is necessary when converting matter density fluctuation fields, defined with respect to the SU local density, to that with respect to the global background density \citep{dePutter:2012}: 
$1+\delta_{\rm m (global)}=(1+\delta_{\rm b}) \times (1+\delta_{\rm m (SU)})$. 
The different definition of the fluctuation fields for matter and halos is motivated by the following observables: weak lensing or cosmic shear arises from matter fluctuation fields with respect to the global mean density, while clustering statistics of large-scale structure tracers, such as galaxies and galaxy clusters, are measured with respect to the local mean density in a survey region \citep{Lietal:14}.
Hence, we compute the $P_{\rm XY}(k)$ response function from equation~(\ref{pkresp_relation}) as:
\begin{align}
 \frac{\mathrm{d} \ln P_{\rm XY}(k)}{\mathrm{d} \delta_{\rm b}} &= n-1 + \left.
 \frac{\mathrm{d} \ln P_{{\rm XY}W}(k_{ W})}{\mathrm{d} \delta_{\rm b}} \right|_k, \nonumber \\
 &\simeq n-1 + \left.
 \frac{\ln P_{{\rm XY}W(+)}(k_{W(+)}) - \ln P_{{\rm XY}W(-)} (k_{W(-)})}{0.02 \times D(z)} \right|_k,
 \label{eq:3D_response}
\end{align}
where $P_{{\rm XY}W(\pm)}$ is measured in the SU simulations with ${\delta_{{\rm b}0}} =\pm 0.01$ at $k_{{W}(\pm)}=(1 \mp |\delta_{\rm b}|/3) \, k$.
The power spectrum has unit of $h^{-3} {\rm Mpc}^3$, not $h_{W}^{-3} {\rm Mpc}^3$.
We numerically evaluate the derivative in the second line. 
Thus, the SU simulation technique provides a computationally inexpensive method to calibrate the {$P(k)$} response. 
In principle, we only need a one-paired simulation to obtain the response if we use identical initial seeds to reduce the sample variance contamination \citep{Lietal:14}. 
This also allows easy computation of the {$P(k)$} response for different cosmological models. 
In this study, we estimate the mean response from $100$-paired SU simulations for the fiducial model. \footnote{How many paired SU simulations are necessary to evaluate the response functions depends on the simulation box size $L$ because the measurement error scales as $L^{-3/2}$. In our case of $L=250 \, h^{-1} {\rm Mpc}$, we use the $100$-paired simulations to obtain accurate results {(see Figure \ref{fig:pk2d_response_errorbar} for the measurement errors)}. If we adopt larger box simulations such as $L\geq 1 \, h^{-1} {\rm Gpc}$, a single pair will be sufficient.}

{Next we evaluate the projected power spectrum response}. 
As described above, the box {sizes} for the SU simulations are identical when converted to the global background. 
For example, for the ${\delta_{{\rm b}0}} =+0.01$ case, $L_W=249.16~h_W^{-1}{\rm Mpc}=249.16\times (h/h_W)~h^{-1}{\rm Mpc}=250~h^{-1}{\rm Mpc}=L$, found in Table~\ref{table1}. 
In other words, we determined the SU simulation box size to satisfy this condition. 
In this study, we consider the same projection thickness as that in the global background to compute the projected density fields. 
Since the 2D power spectrum, $P^{\rm 2D}_{\rm XY (W)}$, relates to the corresponding 3D spectrum, $P_{\rm XY (W)}$, via\footnote{{This relation is derived in Appendix B.}} $P^{\rm 2D}_{\rm XY}= P_{\rm XY}/L$ and $P^{\rm 2D}_{\rm XY W}= P_{\rm XY W}/L$ ($L$ is the width of projection length or the box size),
the same relation found in equation~(\ref{pkresp_relation}) holds for the 2D power spectrum:
\beq 
 P_{\rm XY}^{\rm 2D}(k) \, k^3 = \left( 1+\delta_{\rm b} \right)^n P_{\rm XY W}^{\rm 2D}(k_{\rm W}) \, k_{\rm W}^3.
\label{eq:2D_response}
\eeq
We also obtain the $P^{\rm 2D}_{\rm XY}(k)$ response in a manner that is identical to equation~(\ref{eq:3D_response}) by replacing $P_{\rm XY}$ with $P^{\rm 2D}_{\rm XY}$.
We calculate the mean response from $300$-paired SU samples $( =3 \, {\rm projection~ directions} \times 100 \, {\rm realisations})$.  
Similarly, predictions from perturbation theory, for the 2D response, are obtained by replacing $P_{\rm XY}(k)$ with $P^{\rm 2D}_{\rm XY}(k)$.
Note that the above formula is different from equation~(16) in \citet{TakadaJain:09}, where they employed Limber's approximation, and subsequently performed the azimuthal angle average in the long mode direction to derive the response function.
In this study, we ignore any effects {arising from} large-scale tidal fields on the covariance \citep{Akitsuetal:17,Akitsu:2017b,2017arXiv171107467B}.

\begin{figure}
    \vspace*{-2cm}
	\includegraphics[width=\columnwidth]{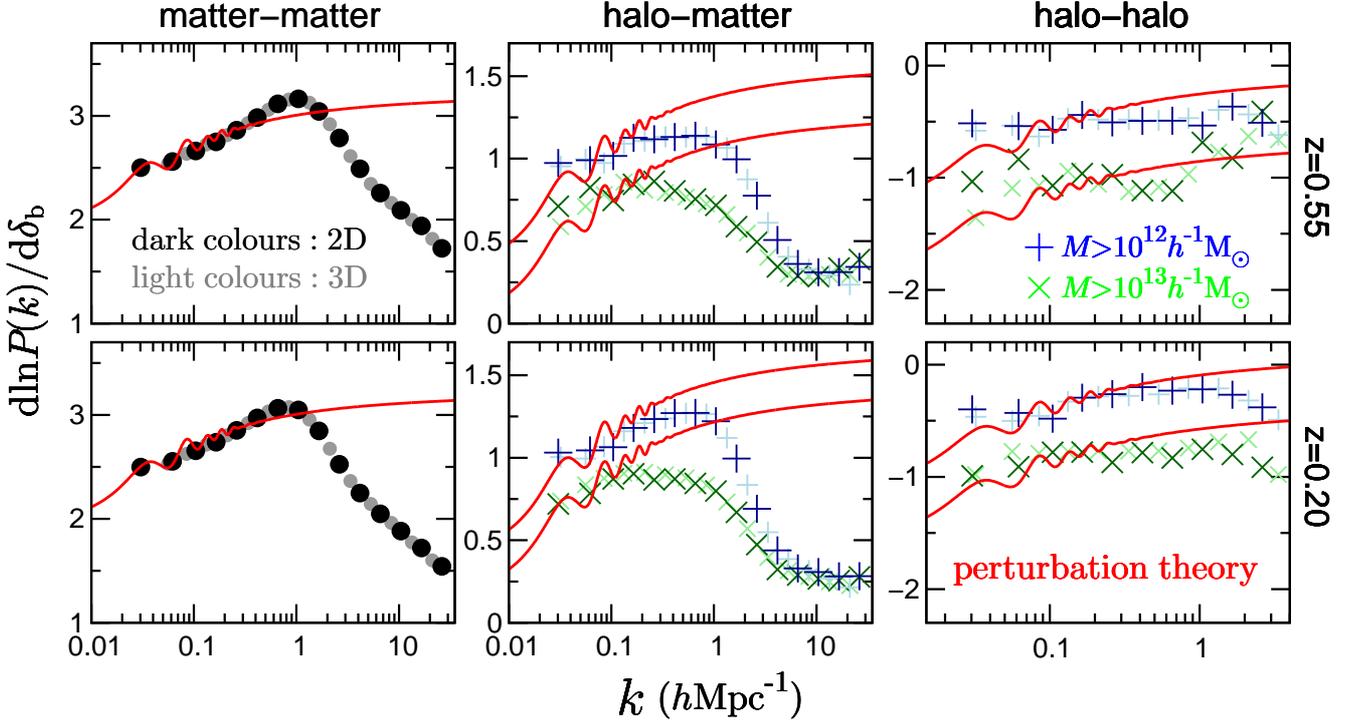}
    \caption{Fractional response functions of power spectra to the large-scale density contrast 
    $\delta_{\rm b}$, $\mathrm{d}\ln P_{\rm XY}(k)/\mathrm{d}\delta_{\rm b}$, for the matter-matter auto-spectrum, matter-halo cross-spectrum and halo-halo auto-spectrum from panels on the left to the right at $z=0.55$ (in the upper row) and $z=0.20$ (in the lower row), respectively. In this fractional form, we expect that the response functions, for the three-dimensional and projected power spectra, are identical: $\mathrm{d}\ln P_{\rm XY}(k)/\mathrm{d}\delta_{\rm b}=\mathrm{d}\ln P^{\rm 2D}_{\rm XY}(k)/\mathrm{d}\delta_{\rm b}$ 
    (see text for details). 
The large dark-coloured symbols in each panel display the response functions of projected power spectra estimated from the SU simulations, while the corresponding small light-coloured symbols are those for the 3D power spectra. 
The plus and cross symbols in the middle and right panels are results for halos with masses $M\!>\!10^{12}$ and $>\!10^{13} \, h^{-1}{\rm M}_\odot$, respectively. 
The solid red curves in each panel are predictions from the perturbation theory (equation~(\ref{pk_resp_pert})), for which we take into account the $k$-binning used in the simulations and use bias parameters, $b_1$ and $b_2$, estimated from halo abundance in the SU simulations (see text for details).
 }
    \label{fig:pk2d_response}
\end{figure}
\begin{figure}
    \vspace*{-2cm}
	\includegraphics[width=\columnwidth]{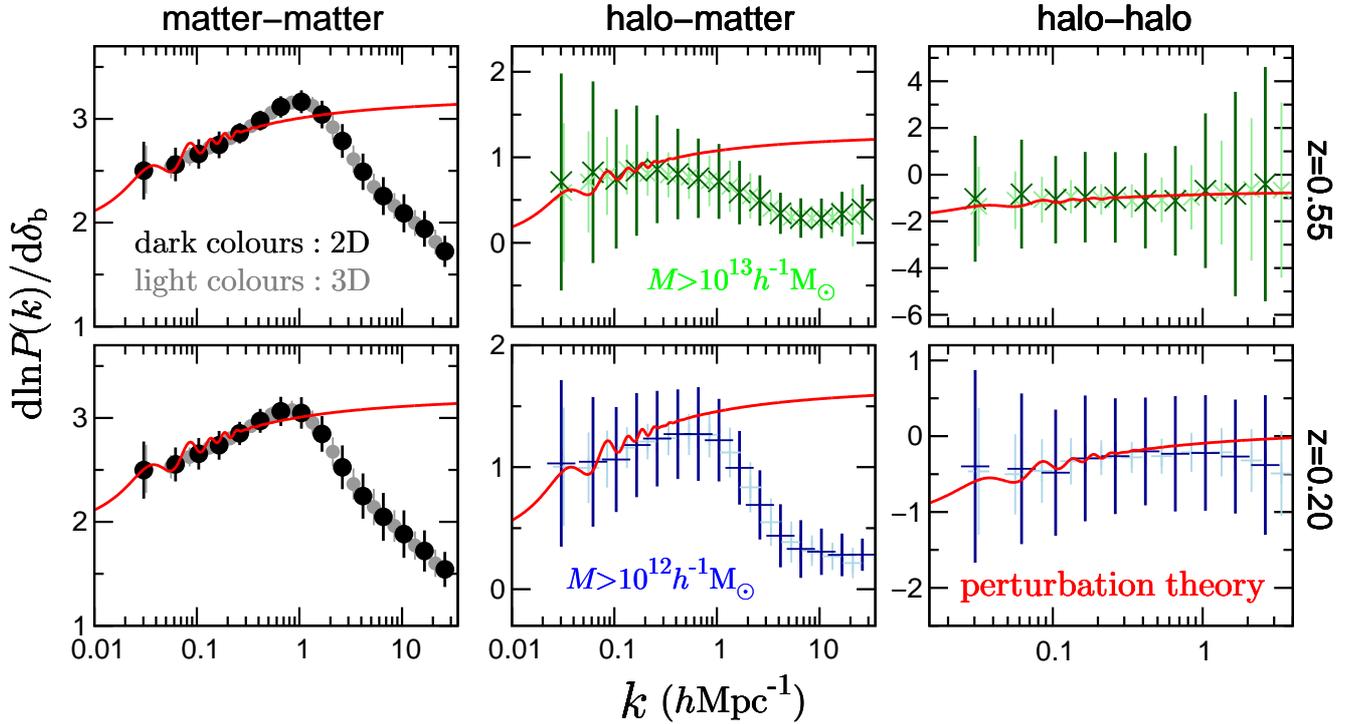}
    \caption{{Same as Figure~\ref{fig:pk2d_response}, but each panel also plots error bars that are estimated from scatters in $100$ realizations. The upper (lower) panels are for $M>10^{13}(10^{12}) \, h^{-1} {\rm M}_\odot$ at $z=0.55 \, (0.20)$.}
 }
    \label{fig:pk2d_response_errorbar}
\end{figure}
\begin{table*}
	\centering
	\begin{tabular}{ccccccc} 
		\hline
	  redshift & $M \, (h^{-1} {\rm M}_\odot)$ & $b_1$ & $b_1$(T10) & $b_2$ & $b_2$(L16) & $b_2$(H17) \\
		\hline
	$0.55$ & $>\!10^{12}$ & $1.30 \pm 0.02$ & $1.22$ & $-0.42 \pm 0.09$ & $-0.72$ & $-0.61$ \\
    $$ & $>\!10^{13}$ & $2.00 \pm 0.05$ & $1.97$ & $0.15 \pm 0.44$ & $0.13$ & $0.14$ \\
    $0.20$ & $>\!10^{12}$ & $1.09 \pm 0.01$ &  $1.02$ & $-0.51 \pm 0.07$ & $-0.74$ & $-0.61$ \\
    $$ & $>\!10^{13}$ & $1.58 \pm 0.03$ & $1.56$ & $-0.33 \pm 0.21$ & $-0.45$ & $-0.39$ \\
		\hline
	\end{tabular}
	\caption{
	The first- and second-order halo bias parameters, $b_1$ and $b_2$, estimated from SU simulations for each halo sample with $M \! > \! 10^{12}$ and $> 10^{13} \, h^{-1} {\rm M}_\odot$ at $z=0.20$ and 0.55, respectively (see text for details). 
	The mean, with error, is evaluated from $10$ SU realisations. 
	The fourth, sixth and seventh columns are from the following previous studies for comparison: \citet{Tinker:2010} (T10) for $b_1$, and \citet{Lazeyras:2016} (L16) and \citet{Hoffmann:2017} (H17) for $b_2$, respectively, where we use the mass function in \citet{Tinker:2008} to compute the mean bias parameters for halo samples at each mass threshold.
	}
	\label{table_bias}
\end{table*}
To validate the simulation results at large scales (small $k$), we compare them with perturbation theory predictions for the 3D fields \citep{Baldaufetal:16}:
\begin{align}
 \frac{\mathrm{d} \ln P_{\rm mm}(k)}{\mathrm{d} \delta_{\rm b}} &= \frac{47}{21} - \frac{1}{3} \frac{\mathrm{d} \ln P_{\rm m,lin}(k)}{\mathrm{d} \ln k},  \nonumber \\
 \frac{\mathrm{d} \ln P_{\rm hm}(k)}{\mathrm{d} \delta_{\rm b}} &= \frac{47}{21} + \frac{b_2}{b_1} - b_1 - \frac{1}{3} \frac{\mathrm{d} 
 \ln P_{\rm m,lin}(k)}{\mathrm{d} \ln k}, 
\label{pk_resp_pert}
\\
 \frac{\mathrm{d} \ln P_{\rm hh}(k)}{\mathrm{d} \delta_{\rm b}} &= \frac{47}{21} + 2 \frac{b_2}{b_1} - 2 b_1 - \frac{1}{3} \frac{\mathrm{d}
  \ln P_{\rm m,lin}(k)}{\mathrm{d} \ln k},
  \nonumber 
\end{align}
where $b_1$ and $b_2$ denote the first- and second-order halo biases. We also use the same SU simulations to estimate these bias parameters \citep{Baldaufetal:16,Lazeyras:2016,Lietal:16}.
For this, we use $10$ paired realisations from extended SU simulations with larger $\delta_{\rm b}$ values, i.e., ${\delta_{{\rm b}0}} = \pm 0.02, \pm 0.04, \pm 0.1,$ and $\pm 0.2$ in addition to $\pm 0.01$, to estimate the nonlinear $b_2$ parameter \citep{Lazeyras:2016,PP:2017}.
{These extra SU simulations are necessary to measure each bias parameter ($b_1$ and $b_2$) separately using equation (\ref{eq:halo_bias}).}
The number of halos heavier than $M$ found in the SU simulation box can be expanded in terms of $\delta_{\rm b}$ with the following equation:
\begin{equation}
 N_{\rm h}(>\! M,z;\delta_{\rm b}) = N_{\rm h}(>\! M,z;\delta_{\rm b}=0) \left[ 1 + b_1^{\rm (L)}(>\! M,z) \, \delta_{\rm b} + 
 \frac{1}{2!} b_2^{\rm (L)}(>\! M,z) \, \delta_{\rm b}^2 + \frac{1}{3!} b_3^{\rm (L)}(>\! M,z) \, \delta_{\rm b}^3 + \cdots \right],
 \label{eq:halo_bias}
\end{equation}
where the coefficients are the Lagrangian biases.
Here, we fit these bias parameters up to the third order.
The Eulerian biases are $b_1=b_1^{\rm (L)}+1$ and $b_2=b_2^{\rm (L)}+(8/21)b_1^{\rm (L)}$ \citep{2002PhR...372....1C,Baldaufetal:16}.
{Therefore, the SU method gives a straightforward way to measure the nonlinear bias by simply counting the halo numbers in the paired SU boxes.}
Table~\ref{table_bias} lists the measured halo biases, $b_1$ and $b_2$, from the $10$ realisations, compared with results from previous studies.\footnote{
Based on previous studies, the mean halo biases for halos heavier than $M$ {are} calculated as follows:
\begin{equation}
 b_1(> \!\! M,z) = \frac{1}{n_{\rm h}(>\!\!M,z)} \int_M^{\infty} \! \mathrm{d}M^\prime \frac{\mathrm{d}n}{\mathrm{d}M}(M^\prime,z) \, b_1(M^\prime,z),    
 ~b_2(> \!\! M,z) = \frac{1}{n_{\rm h}(>\!\!M,z)} \int_M^{\infty} \! \mathrm{d}M^\prime \frac{\mathrm{d}n}{\mathrm{d}M}(M^\prime,z) \, b_2[b_1(M^\prime,z)], 
\nonumber
\end{equation}
where $n_{\rm h}(>\!\!M,z)=\int_M^{\infty} \mathrm{d}M^\prime [\mathrm{d}n/\mathrm{d}M](M^\prime,z)$ is the cumulative halo number density. 
We used the fitting functions from \cite{Tinker:2008,Tinker:2010} for the halo mass function and linear halo bias as well as functions from \cite{Lazeyras:2016} and \cite{Hoffmann:2017} for the second-order halo bias, $b_2=b_2(b_1)$. 
}
Our results for $b_1$ are consistent with those of \cite{Tinker:2010} within $10 \%$, while our results for $b_2$, which are noisy, agree well with those of \cite{Lazeyras:2016} and \cite{Hoffmann:2017} within $|\Delta b_2|=0.2$. \cite{Hoffmann:2017} has better agreement but predicts slightly smaller values than those in this study.
The $b_2$ for halos with $M \!>\! 10^{12} \, h^{-1} {\rm M}_\odot$ at $z=0.55$ exhibits a sizable difference from previous studies, but the reason for this is beyond the scope of this study.

Figure \ref{fig:pk2d_response} shows the response functions for matter-matter (left panel), halo-matter (middle), and halo-halo (right) spectra.
First, response functions have large amplitudes, with an order of matter auto-, matter-halo cross- and halo auto-spectra. 
Large amplitudes for matter spectra compared with the halo spectra are due to the use of the global mean mass density in the definition of the mass
density fluctuation field (the term of $n-1$ in equation~(\ref{eq:3D_response})). 
The large dark-coloured symbols in each panel are the response functions for the 2D power spectra, while the corresponding small light-coloured symbols are for the 3D spectra. 
The comparison clearly shows that response functions for the 2D and 3D spectra agree with each other or are identical to within scatters in both the linear and nonlinear regimes, validating the arguments near equation~(\ref{eq:2D_response}). 
Simulation results are consistent with theoretical predictions in the linear regime {up to $k=0.2 \, h {\rm Mpc}^{-1}$ within $3 \, (10) \, \%$ for $P_{\rm mm}$ ($P_{\rm hm }$ and $P_{\rm hh}$)}. 
Note that, for perturbation theory predictions, we take into account $k$-binning used in the simulation results; we smooth out perturbation theory predictions within a given $k$-bin, which smears baryon acoustic oscillation (BAO) features and provides better matches to the simulation results.

{Figure \ref{fig:pk2d_response_errorbar} plots the measurement errors of $P(k)$ responses from $100$ realizations. The 2D errors are always larger than the 3D errors because the number of mode in the $k$ bin, $N_k$, is smaller (or, in other words, the information of density contrast in the projection direction is lost in the 2D case). For the halo-halo component, the heavier halo sample has larger errors at small scale because of larger shot noise. The errors are smaller with an order of $P_{\rm mm}$, $P_{\rm hm}$, and $P_{\rm hh}$.}

\subsection{Covariances of the power spectra}
\label{sec:covariance}

\begin{table*}
	\centering
	\begin{tabular}{cccc} 
		\hline
	 & $L \, (h^{-1} {\rm Mpc})$ & $N_{\rm p}^3$ & $N_{\rm r}$ \\
		\hline
	small box & $250$ & $512^3$ & $100$ \\
    large box & $1000$ & $2048^3$ & $20$ \\
		\hline
	\end{tabular}
	\caption{$N$-body simulation parameters used to test and validate the covariance matrix of {$P^{\rm 2D}(k)$} for matter-matter, matter-halo and halo-halo spectra. To do this, we use two different box-size $N$-body simulations, i.e., small-box and large-box simulations, whereas spatial resolution remains the same. 
	Each column provides the simulation box size ($L$), the number of particles ($N_{\rm p}^3$), and the number of realisations ($N_{\rm r}$), respectively. We use simulations at two redshifts, $z=0.20$ and $0.55$.}
	\label{table2}
\end{table*}

\begin{figure}
    \begin{center}
	\includegraphics[width=0.6\columnwidth]{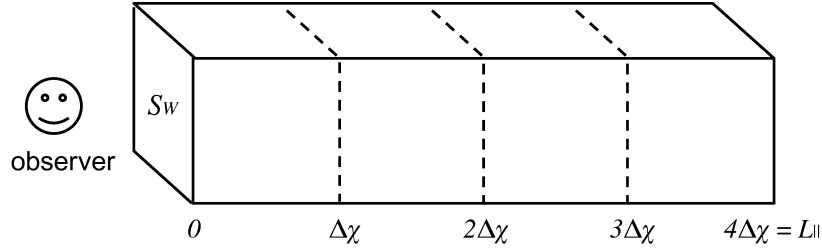}
    \end{center}
    \caption{An illustration of the configuration of the $N$-body simulation boxes used for {mock} experiments of {$P^{\rm 2D}(k)$} measurements. We consider that $\Delta \chi = 250 \, h^{-1}{\rm Mpc}$ for a subvolume thickness used to define the projected field in each plane, $L_\parallel=1000 \, h^{-1}{\rm Mpc}$ for the total projection thickness and $S_W=250^2 \, (h^{-1}{\rm Mpc})^2$ for the square-shaped survey area in the flat plane perpendicular to the line-of-sight direction. For simplicity, we ignore redshift evolution; we use the $N$-body data at the same redshift, $z=0.20$ or $0.55$. 
    }
    \label{fig:boxes_in_line}
\end{figure}

\begin{table*}
  \centering
  \begin{tabular}{|l|l|l|l|l|} \hline
      Name & SSC term & Projection length & Perpendicular \& parallel  & Characteristics \\
           &   (equation)     &    ($h^{-1} \, {\rm Mpc}$)  & super-survey modes &  \\ 
      \hline \hline
      Case (i) & (\ref{eq:SSC_caseI}) & $1000$ & Included \& No & Use $250^2\times 1000 \, (h^{-1}{\rm Mpc})^3$ rectangular box for the projection. \\ 
      &&&&    \\
      \hline
      Case (iia) & (\ref{eq:SSC_caseII}) & $4 \times 250$ & Included \& No & Use the same rectangular box in Case~(i),  \\
        &&&& but divide it into four cubic boxes before projection.
      \\
      \hline
      Case (iib) & (\ref{eq:SSC_caseII}) & $4 \times 250$ & Included \& Included & Similar to Case~(iia), but use four cubic boxes randomly taken \\
        &&&& from different large-box realisations. \\
      \hline
      Case (iii) & No & $4 \times 250$ & No \& No & Similar to Case~(iia,b), but use four cubic small-boxes \\
&&&&   with periodic boundary conditions. \\
      \hline
    \end{tabular}
    \caption{A summary of our numerical experiments in Section~\ref{sec:covariance}. All the cases use the simulations, each of which has
    a total volume
    of $250^2 \times 1000 \, (h^{-1}{\rm Mpc})^3$, to generate the projected fields, where 
    $250^2 \, (h^{-1}{\rm Mpc})^2$ is the projected area (see Fig.~\ref{fig:boxes_in_line}). 
    }
  \label{table3}
\end{table*} 

\begin{figure}
	\includegraphics[width=\columnwidth]{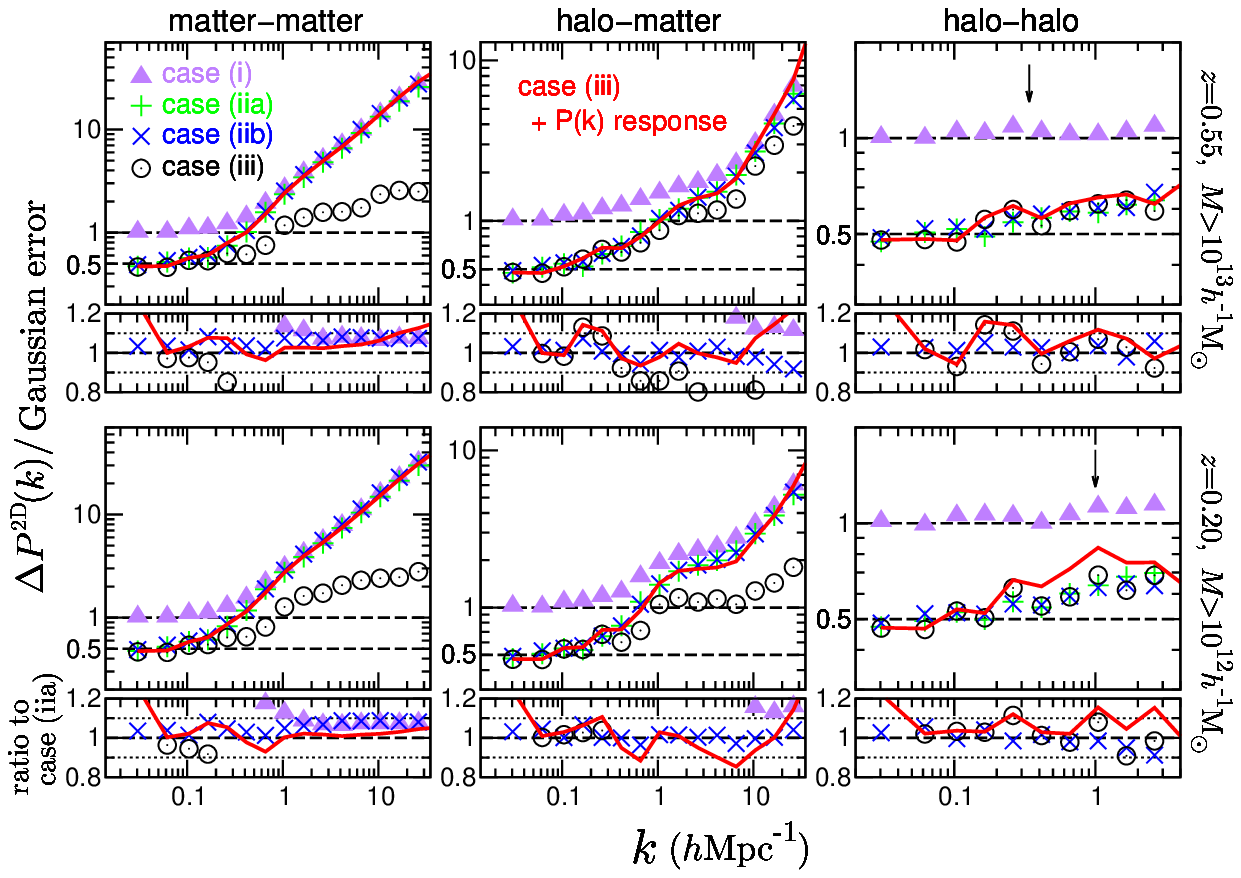}
    \vspace*{0cm}
    \caption{The square root of the variance of projected power spectra for matter-matter, matter-halo, and halo-halo from the left to right panels, respectively. 
    The variances are estimated from the light-cone realisations of $N$-body simulations that have total volumes of $250^2\times 1000 \, (h^{-1}{\rm Mpc})^3$, as illustrated in Fig.~\ref{fig:boxes_in_line}. 
    By doing this, we consider the following four different cases: Case~(i) is a projection of the whole rectangular volume, and Cases~(iia,b)-(iii) are from superpositions of the four small-boxes, each of which has a volume of $250^3 \, (h^{-1}{\rm Mpc})^3$. Case~(iia) has contributions from super-survey modes in the perpendicular direction, while Case~(iib) has super-survey modes in both parallel and perpendicular directions. Case~(iii) has small boxes with periodic boundary conditions (i.e., no super-survey mode effects). 
    We show all the results relative to expectations from the Gaussian {(G)} variance for Case~(i). We expect that the {G} variance for Cases~(iia,b) and (iii) are smaller than that for Case~(i) by a factor of $2$, as denoted by the two horizontal dashed lines.
    The solid red curves in each panel are the results obtained by adding the SSC contribution to Case~(iii) (the circle symbols), which efficiently reproduces the results of Cases~(iia,b).
    The upper (lower) panels show results for halos with $M>10^{12} \, (10^{13}) \, h^{-1}{\rm M}_\odot$ at $z=0.55 \, (0.20)$. The black arrows, in the right panels, indicate scales where $P^{\rm 2D}_{\rm hh}(k)$ equals the shot noise. The $k$-bin width, which the {G} error expectation depends on, is $\Delta \log k = 0.2$.
    {The small bottom panels plot ratios to Case (iia). The horizontal dotted lines denote $\pm 10 \%$ difference.}}
    \label{fig:pk2d_variance}
\end{figure}
In this section, we analyse the covariance matrices of projected power spectra, $P^{\rm 2D}(k)$, using a sufficient number of $N$-body realisations. 
We ran the following two types of simulations for the fiducial cosmological model (as given in the first line of Table~\ref{table1}): simulations with small boxes, side length $L=250 \, h^{-1}$Mpc with $512^3$ particles, and those with large boxes, $L=1000 \, h^{-1}$Mpc with $2048^3$ particles, summarised in Table \ref{table2}. 
The number of realisations performed is $100$ and $20$ for the small and large boxes, respectively.
We divide each large-box simulation into $4^3=64$ cubic subvolumes with side lengths of $250 \, h^{-1}{\rm Mpc}$, {translating into} $1280$ subvolumes in total. 
To study the impact of super-survey modes on {$P^{\rm 2D}(k)$}, we perform the following numerical experiments using simulations constructed from different combinations of  the large-box, small-box and subvolume realisations. 
{The experiments will reveal the SSC effect in the perpendicular direction (Case (i)) and in the perpendicular and parallel directions (Case (ii)) by comparing with no SSC case (Case (iii)).}
In all of the below Cases (i)-(iii), we use a survey area of $S_{\rm W}=250^2 \, (h^{-1} {\rm Mpc})^2$ and a projection length of $L_\parallel=1000 \, h^{-1} {\rm Mpc}$ (see also the simulation box configuration in Fig. \ref{fig:boxes_in_line}).

\begin{itemize}
\item{\bf Case~(i)} We use a rectangular-shaped subvolume of $250^2 \times 1000 \, (h^{-1} {\rm Mpc})^3$ cut out of a large-box realisation to create projected matter and halo fields {on} square-shaped areas of $250^2 \, (h^{-1} {\rm Mpc})^2$. We use a direction length of $1000\, h^{-1}{\rm Mpc}$ for the projection (line-of-sight) direction.
Then, we compute {$P^{\rm 2D}(k)$} from the two-dimensional data by FFT. 
The power spectra measured in this manner include the effects of super-survey modes perpendicular to the line-of-sight. 
Note that, in this case, the density fluctuation field is continuous and obeys periodic boundary conditions in the line-of-sight direction. 
Since we generate the projected field from the $N$-body simulations, at the same redshift output, the density fields have no redshift (line-of-sight) evolution in a statistical sense, and the power spectrum of the projected fields is simply
\begin{equation}
P^{\rm 2D}_{\rm XY}(k)=\frac{1}{L_\parallel}P_{\rm XY}(k),
\end{equation}
where $L_\parallel=1000 \, (h^{-1}{\rm Mpc})$ is the projection width and $P_{\rm XY}(k)$ is the 3D power spectrum. 
This is found by setting $F_{\rm X}(\chi)=F_{\rm Y}(\chi)=1/L_\parallel$ in equation~(\ref{eq:pk2d_def}).
\item{\bf Case~(iia)} The same rectangular-shaped volumes as those in Case~(i) are used. In this case, however, we first divide each realisation into four cubic subvolumes of $250^3\,(h^{-1}{\rm Mpc})^3$ each and then project the field along the line-of-sight direction in each subvolume. Then, we estimate {$P^{\rm 2D}(k)$} in each subvolume and finally average {them} to estimate {$P^{\rm 2D}(k)$} for the original rectangular volume, as shown in Fig.~\ref{fig:boxes_in_line}.  
Similarly, by setting $F_{\rm X}(\chi)=F_{\rm X}(\chi)=1/L_\parallel$ in equation~(\ref{eq:pk2d_def}), we can express the {$P^{\rm 2D}(k)$} estimator using a discrete summation as 
\begin{equation}
\widehat{P}^{\rm 2D}_{\rm XY}(k)= \frac{(\Delta \chi)^2}{(L_\parallel)^2}\sum_{i=1}^{N_{\rm sub}} \, \widehat{P}^{\rm 2D}_{{\rm XY}(i)}(k) =\frac{(\Delta \chi)^2}{(L_\parallel)^2}\sum_{i=1}^{N_{\rm sub}} \, \widehat{P}^{\rm 2D}_{{\rm XY}(i)}(k;\delta_{{\rm b}i}), 
\label{eq:caseII}
\end{equation}
where $\Delta\!\chi=250 \, h^{-1}{\rm Mpc}$ is the width of each subvolume, and $\widehat{P}^{\rm 2D}_{{\rm XY}(i)}(k)$ is the projected power spectrum estimator in the $i$-th subvolume. 
The number of subvolumes is $N_{\rm sub}=L_\parallel/\Delta \chi=4$.
As mentioned previously, the local mean density of halos in the survey area is used to estimate the halo power spectra ($\widehat{P}^{\rm 2D}_{{\rm hm}}(k)$ and $\widehat{P}^{\rm 2D}_{{\rm hh}}(k)$).
In the above equation, we explicitly denote that the power spectrum, in each subvolume, has a modulation due to the super-survey mode, $\delta_{{\rm b}i}$.
Similarly, the ensemble average of the above estimator is identical to Case~(i):
\begin{equation}
\avrg{\widehat{P}^{\rm 2D}_{\rm XY}(k)}= \frac{(\Delta \chi)^2}{(L_\parallel)^2}\sum_{i=1}^{N_{\rm sub}} \, \avrg{\widehat{P}^{\rm 2D}_{{\rm XY}(i)}(k)}
= \frac{(\Delta \chi)^2}{(L_\parallel)^2} \frac{N_{\rm sub}}{\Delta \chi} P_{\rm XY}(k) =\frac{1}{L_\parallel}P_{\rm XY}(k),
\end{equation}
where we used $\avrg{\widehat{P}^{\rm 2D}_{{\rm XY}(i)}(k)} = P_{\rm XY}(k)/\Delta \chi$.
Note that this case preserves a continuous radial mode in each realisation. 
\item{\bf Case~(iib)} This case is similar to Case~(iia), but we randomly choose four cubic subvolumes of $250^3 \, (h^{-1}{\rm Mpc})^3$ each, which are taken from the different large-box simulations to construct each rectangular-shaped realisation of $250^2\times 1000 \, (h^{-1}{\rm Mpc})^3$.
These are placed into the four subvolumes along the line-of-sight direction (as illustrated in Fig.~\ref{fig:boxes_in_line}). 
Then, we estimate {$P^{\rm 2D}(k)$} in the same manner as Case~(iia).
This case includes super-survey modes over subvolumes in both perpendicular and parallel directions to the line-of-sight, and the density fluctuations are discontinuous between different subvolumes. Comparing the results of Cases~(i), (iia) and (iib) reveals a radial super-survey mode effect. 

\item{\bf Case~(iii)} This is similar to Cases~(iia,b), but we use four small-box realisations, each of which has a volume of $250^3 \, (h^{-1}{\rm Mpc})^3$, obeying the periodic boundary conditions, to construct rectangular-shaped realisations of $250^2\times 1000 \, (h^{-1}{\rm Mpc})^3$.
This case does not include super-survey modes {neither} perpendicular {nor} parallel to the line-of-sight (no modes beyond the small-box size). 
\end{itemize}
We briefly summarise the characteristics of the above cases in Table \ref{table3}.  
Below, we use $960$ realisations (=$4^2$ patches $\times$ $3$ projections $\times$ $20$ large-box realisations) for Cases~(i)-(iia,b), where we use $4^2=[1000 \, h^{-1}{\rm Mpc}/(250 \, h^{-1}{\rm Mpc})]^2$ patches for each $x$, $y$ or $z$-direction projection, respectively. 
For Case~(iii) we use $75$ realisations (=$100/4$ subsets of 100 small-box realisations $\times$ $3$ projections).
{Measurement error of covariance estimated from $N_{\rm r}$ realizations is $\simeq (N_{\rm r}/2)^{-1/2}$, resulting in $0.45 \, \%$ and $16 \, \%$ for Cases (i-ii) and Case (iii), respectively.}
Here, we show {our} results at $z=0.20$ and $0.55$. As described above, the ensemble average of the {$P^{\rm 2D}(k)$} estimator should be identical for all Cases~(i)-(iii) (or designed to satisfy this condition). 
Then, we use these realisations to estimate the {$P^{\rm 2D}(k)$} covariance matrices for all these cases. 

We discuss theoretical expectations for the covariance matrices of Cases~(i)-(iii). 
First, we consider the Gaussian {(G)} covariance contribution. 
Since we estimate {$P^{\rm 2D}(k)$} from four cubic subvolumes for Cases~(ii)-(iii), we obtain a different {G} covariance contribution compared with Case~(i), which estimates {$P^{\rm 2D}(k)$} for the whole rectangular volume. 
From equation~(\ref{eq:caseII}), we can derive the {G} covariance matrix for Cases~(ii)-(iii):
\begin{align}
{\rm Cov}^{\rm G}_{\rm XY,(ii)-(iii)}(k,k')&=\frac{(\Delta \chi)^4}{(L_\parallel)^4}\frac{1}{N_{\rm mode}(k)}\sum_{i=1}^{N_{\rm sub}}
\left[ \left\{ P^{\rm 2D}_{{\rm XY}(i)}(k) \right\}^2 + P^{\rm 2D}_{{\rm XX}(i)}(k) P^{\rm 2D}_{{\rm YY}(i)}(k)\right]\delta^K_{kk'}\nonumber\\
&=\frac{(\Delta \chi)^2}{(L_\parallel)^2}\frac{N_{\rm sub}}{N_{\rm mode}(k)} \left[ \left\{ P^{\rm 2D}_{\rm XY}(k) \right\}^2 +P^{\rm 2D}_{\rm XX}(k)P^{\rm 2D}_{\rm YY}(k)\right]\delta^K_{kk'}\nonumber\\
&= \frac{1}{N_{\rm sub} N_{\rm mode}(k)} \left[ \left\{ P^{\rm 2D}_{\rm XY}(k) \right\}^2 +P^{\rm 2D}_{\rm XX}(k)P^{\rm 2D}_{\rm YY}(k)\right]\delta^K_{kk'}\nonumber\\
&=\frac{1}{N_{\rm sub}}{\rm Cov}^{\rm G}_{\rm XY,(i)}(k,k'), 
\label{eq:CovG_caseII}
\end{align}
where we used $P^{\rm 2D}_{{\rm XY}(i)}(k)=(L_\parallel/\Delta \chi) P^{\rm 2D}_{\rm XY}(k) = 4 P^{\rm 2D}_{\rm XY}(k)$ and we assumed that the density fields, in different slices (i.e., subvolumes), are independent of each other (see Fig.~\ref{fig:boxes_in_line}). 
$N_{\rm mode}(k)$ is the number of modes used to estimate {$P^{\rm 2D}(k)$} at the $k$-bin, $N_{\rm mode}(k)\simeq k \Delta k S_W /(2\pi)$ for $k \gg S_W^{-1/2}$. 
In the last line, ${\rm Cov}^{\rm G}_{\rm XY,(i)}$ is the {G} covariance for Case~(i), showing that the {G} covariance for Cases~(ii)--(iii) is smaller in amplitude than that of Case~(i) by a factor of $N_{\rm sub}=4$. 
 
Next, we consider SSC contribution that affects the results for Cases~(i)--(ii). By inserting $F_{\rm X}(\chi)=F_{\rm Y}(\chi)=1/L_\parallel$ into equation~(\ref{eq:ssc_derivation}), the SSC contribution for Case~(i) is 
\begin{equation}
{\rm Cov}^{\rm SSC}_{\rm XY,(i)}(k,k')
= \frac{\partial P^{\rm 2D}_{\rm XY}(k)}{\partial \delta_{\rm b}}
\frac{\partial P^{\rm 2D}_{\rm XY}(k')}{\partial \delta_{\rm b}}\sigma_{\rm b}^2(V_{\rm tot}), \label{eq:SSC_caseI}
\end{equation}
where $V_{\rm tot}= 250^2 \times 1000 \, (h^{-1}{\rm Mpc})^3$, and we use the fact that the density field does not evolve along the projection direction.
Similarly, we solve for the SSC covariance for Cases~(iia) and (iib):
\begin{align}
{\rm Cov}^{\rm SSC}_{\rm XY,(ii)}(k,k')&=\frac{(\Delta \chi)^4}{(L_\parallel)^4}\sum_{i=1}^{N_{\rm sub}} \frac{\partial P^{\rm 2D}_{{\rm XY}(i)}(k)}{\partial \delta_{\rm b}} \frac{\partial P^{\rm 2D}_{{\rm XY}(i)}(k')}{\partial \delta_{\rm b}}\sigma_{\rm b}^2(V_{\rm sub})\nonumber\\
&= \frac{(\Delta \chi)^2}{(L_\parallel)^2} N_{\rm sub}
\frac{\partial P^{\rm 2D}_{\rm XY}(k)}{\partial \delta_{\rm b}}
\frac{\partial P^{\rm 2D}_{\rm XY}(k')}{\partial \delta_{\rm b}}\sigma_{\rm b}^2(V_{\rm sub}), \nonumber\\
&\simeq  
\frac{\partial P^{\rm 2D}_{\rm XY}(k)}{\partial \delta_{\rm b}}
\frac{\partial P^{\rm 2D}_{\rm XY}(k')}{\partial \delta_{\rm b}}\frac{1}{N_{\rm sub}}\sigma_{\rm b}^2(V_{\rm sub}) \nonumber \\
&= {\rm Cov}^{\rm SSC}_{\rm XY,(i)}(k,k'),
\label{eq:SSC_caseII}
\end{align}	
where $V_{\rm sub}=250^3 \, (h^{-1}{\rm Mpc})^3$, and the last equality originates from the following equality for large-scale density variance:
\begin{equation}
\sigma_{\rm b}^2(V_{\rm tot})\simeq \frac{1}{N_{\rm sub}}\sigma_{\rm b}^2(V_{\rm sub}). 
\end{equation}
The above relation was checked for validity, by comparing the variances of rectangular volume with those of subvolume. 
We expect that SSC covariance is similar to Cases~(i)--(ii), which we will verify below with simulations.

Fig.~\ref{fig:pk2d_variance} displays the square root of the variance (i.e., the diagonal covariance elements) for the matter-matter, matter-halo and halo-halo {$P^{\rm 2D}(k)$}. 
For illustrative purposes, we show the variance relative to the {G} error of Case~(i). 
The triangle, plus, cross and circle symbols, in each panel, show the variances measured from simulations for Cases~(i), (iia), (iib), and (iii), respectively. 
All results are near unity or $0.5$ at $k\simlt 0.2~h/{\rm Mpc}$, i.e., consistent with {G} error expectations, where the {G} error for Cases~(iia,b) and (iii) should be smaller than that of Case~(i) by a factor of $2$, as shown by equation~(\ref{eq:CovG_caseII}). 
On the other hand, non-Gaussian (NG) contributions greatly exceed {G} error for {$P^{\rm 2D}_{\rm mm}(k)$ and $P^{\rm 2D}_{\rm hm}(k)$} in the nonlinear regime of $k\simgt 0.2~h/{\rm Mpc}$, while {NG} error for {$P^{\rm 2D}_{\rm hh}(k)$} does not appear to be significant because shot noise is dominant at such small scales.
Comparing the results for Cases~(iia,b) with Case~(iii), we observe that SSC provides a dominant contribution to {NG} errors, while results that deviate from $0.5$ for Case~(iii) show the contribution of {\it connected} {NG} error from equation~(\ref{eq:cov_def}) because Case~(iii) uses simulations with periodic boundary conditions (i.e., there are no super-survey modes beyond the area in the projected plane). 
The solid red curve in each panel shows results obtained by summing the variance for Case~(iii) and the SSC term (equation~(\ref{eq:SSC_caseII})), where we used response functions measured from SU simulations (in Section \ref{sec:response}), and directly estimated $\sigma_{\rm b}$ from rectangular realisations (from variations of $N$-body particles in the realisations). 
The solid curves agree remarkably with results for Case~(iia,b) over the wide range of scales as well as results for Case~(i) in the nonlinear regime {within $10 \, \%$ accuracy (see rations in the small bottom panels)}. 
From these results, we observe no strong evidence of large-scale tidal field effects, which we ignored in the SSC calibration method. This does not seem consistent with results from \cite{2017arXiv171107467B}, who claimed that large-scale tidal fields cause a {small ($\sim 5 \, \%$)} additional contribution to the SSC term in cosmic shear covariance (which corresponds 
to the results for matter-matter covariance in Fig.~\ref{fig:pk2d_variance}).
{However, our results are still noisy to validate their tidal-field contribution, and more $N-$body realizations are necessary to reach a confident conclusion.}   
{Larger} SSC contributions for {$P^{\rm 2D}_{\rm mm}(k)$} relative to {$P^{\rm 2D}_{\rm hm}(k)$} or for {$P^{\rm 2D}_{\rm hm}(k)$} relative to {$P^{\rm 2D}_{\rm hh}(k)$} are due to {larger} amplitudes of response functions, as shown in Fig.~\ref{fig:pk2d_response}. Furthermore, results for Case~(iib) do not show any sizable {(less than $10 \, \%$)} deviation from Case~(iia) (or Case~(i) in the nonlinear regime), indicating that the large-scale mode parallel to the projection direction has a negligible impact on the covariance.

\begin{figure}
	\includegraphics[width=1.\columnwidth]{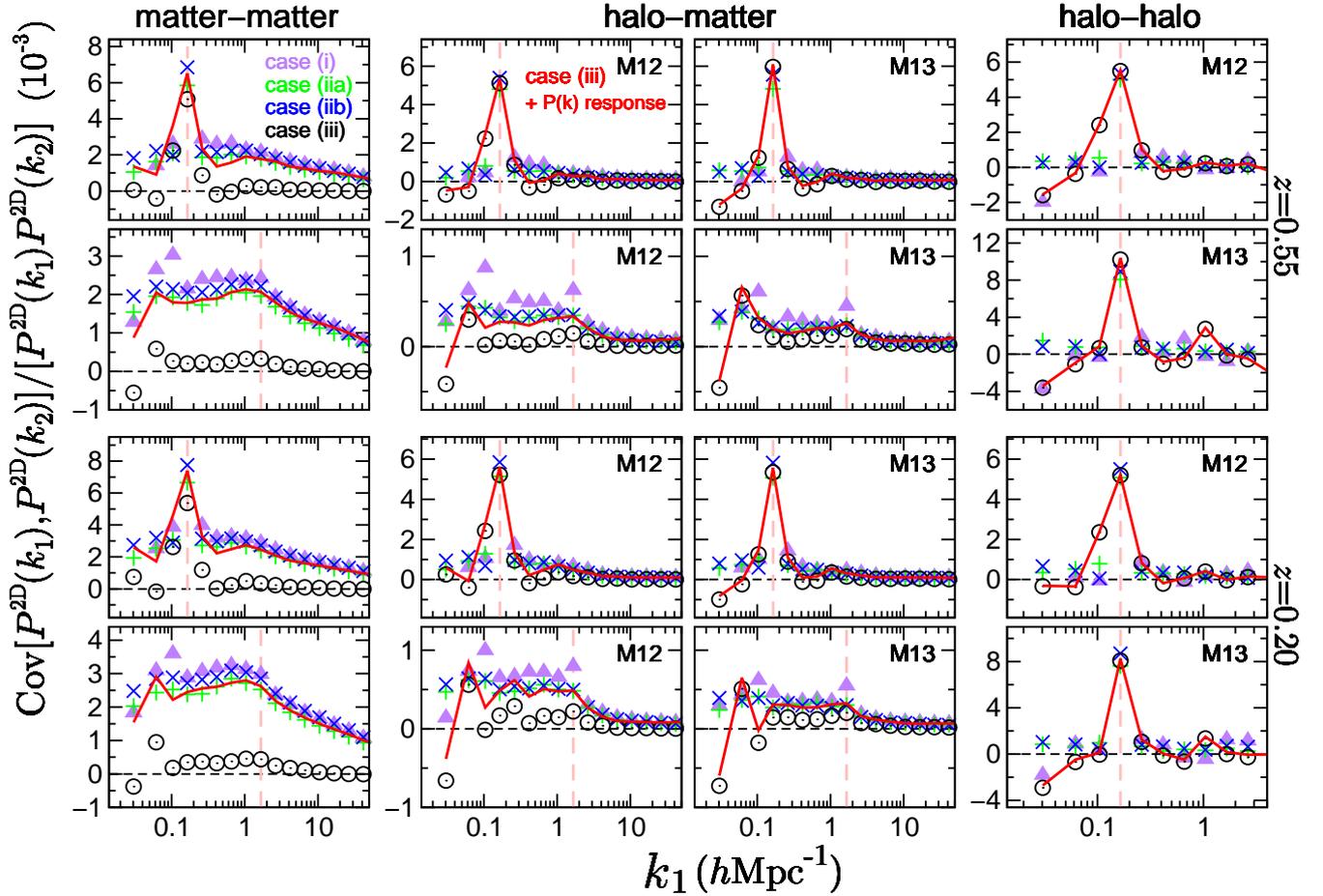}
	\vspace*{-1em}
    \caption{Similar to Fig.~\ref{fig:pk2d_variance}. However, here we show the off-diagonal elements of the covariance matrices. For illustrative purposes, we display the ${\rm Cov}[P^{\rm 2D}_{\rm XY}(k_1),P^{\rm 2D}_{\rm XY}(k_2)]/[P^{\rm 2D}_{\rm XY}(k_1)P^{\rm 2D}_{\rm XY}(k_2)]$ in units of $10^{-3}$ for a fixed $k_2$ value with varying $k_1$ along the $x$-axis. The vertical dashed {pink} line in each panel denotes the chosen $k_2$ value. The upper two-row panels show results at $z=0.55$, while the lower two-row panels are at $z=0.20$. Panels labelled as M13 and M12 are for halo samples with $M \!>\! 10^{13}$ and $> \! 10^{12} \, h^{-1} {\rm M}_\odot$, respectively. The solid red curve in each panel shows the SSC response added to Case~(iii). 
}
    \label{fig:pk2d_covariance}
\end{figure}
Fig.~\ref{fig:pk2d_covariance} displays results that are similar to Fig.~\ref{fig:pk2d_variance}; however, these results are for off-diagonal covariance components. 
This figure confirms the results shown in Fig.~\ref{fig:pk2d_variance}; the {NG} and SSC contributions are significant in the nonlinear 
regime, especially for {$P^{\rm 2D}_{\rm mm}(k)$} or {$P^{\rm 2D}_{\rm hm}(k)$}. The results (circle symbols) for Case~(iii) (periodic boundary condition) 
show a deviation from Cases~(i) and (iia,b) at $k\simlt 0.1~h/{\rm Mpc}$ in the linear regime due to the effect of window function; the cutout of $(250)^2~(h^{-1}{\rm Mpc})^2$ patches from larger-area simulation in Cases~(i) and (iia,b) modifies {$P^{\rm 2D}(k)$} at small {$k$-modes}.

\section{Comparison with ray-tracing simulation results}
\label{sec:full_sky_simulations}

So far, we have studied the response functions and SSC contributions for matter-matter, matter-halo and halo-halo projected power spectra using {cosmological simulation cubical volumes}. 
In this section, we develop a method to calibrate covariances for cosmic shear and galaxy--galaxy lensing by combining the model ingredients that we have computed.
{We further consider the covariances of configuration-space correlation functions, instead of power spectra, because the former} are more commonly used in actual measurements. 
We first derive the response functions of the correlation functions to $\delta_{\rm b}$, and subsequently test the method by comparing the covariances estimated based on this method, with those estimated using light-cone ray-tracing simulations of weak lensing and halo fields. 

\subsection{Response of 2D correlation function}
\label{sec:2dxi_response}

\begin{figure}
    \vspace*{-2cm}
	\includegraphics[width=\columnwidth]{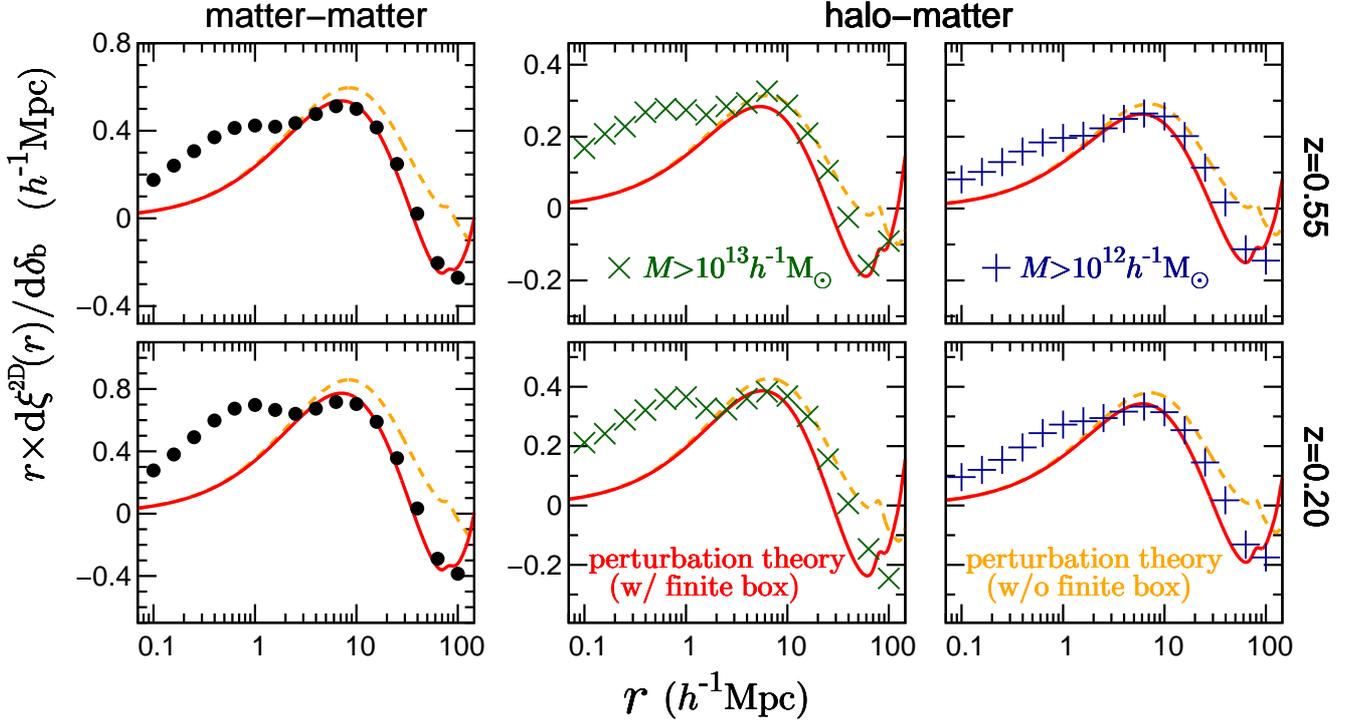}
	\vspace*{-2em}
    \caption{Responses of the projected correlation functions to large-scale density, $\delta_{\rm b}$, for matter-matter and halo-matter at $z=0.55$ (upper panel) and $z=0.20$ (lower), respectively. Here, $r\times \mathrm{d}\xi^{\rm 2D}_{\rm XY}(r)/\mathrm{d}\delta_{\rm b}$ is plotted for illustrative purposes. For the matter-halo response, we show the results for halos with $M>10^{13}$ or $10^{12} \, h^{-1}M_\odot$ in the middle and right panels, respectively. 
    The circle, cross and plus symbols are results obtained from 2D Fourier transformations of the 2D power spectrum responses estimated from the SU simulations in Fig.~\ref{fig:pk2d_response}. 
    For comparison, the solid-red curves show the perturbation theory predictions (equation~\ref{xi2d_resp_pert}) where we properly employ the minimum wavenumber $k_{\rm min}=2\pi/L$ with $L=250~h^{-1}{\rm Mpc}$, which is the box size of SU simulations, for the $k$-integration (labelled as ``finite box''). The dashed-orange curves are the results for $k_{\rm min}=0$. 
	}
    \label{fig:xi2d_response}
\end{figure}
\begin{figure}
    \vspace*{-2cm}
	\includegraphics[width=\columnwidth]{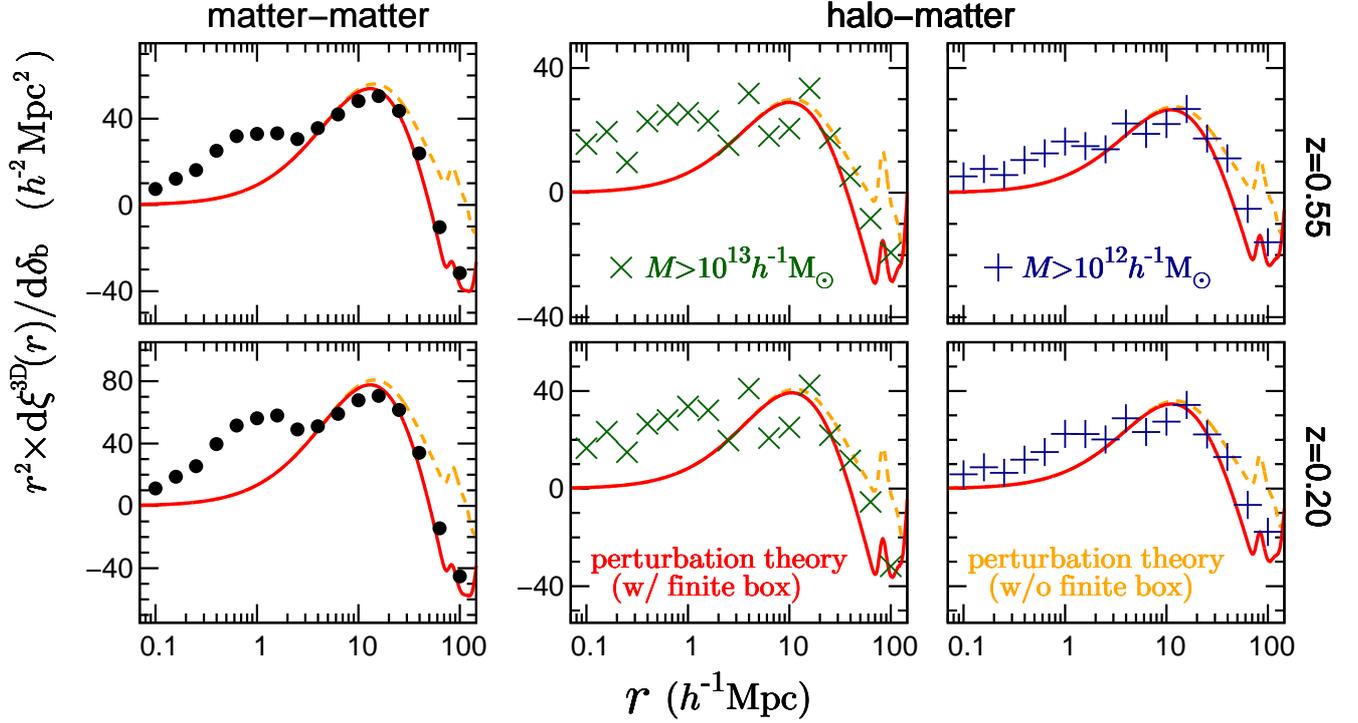}
    \vspace*{-2em}
    \caption{Similar to Fig.~\ref{fig:xi2d_response}, but showing the responses of the 3D correlation functions.
	 Comparing this with the previous figure indicates that the amplitude of the 2D response is smaller than that of the 3D response by $\xi^{\rm 2D}(r) \propto L^{-1}$, where $L$ is the projection width. 
    }
    \label{fig:xi3d_response}
\end{figure}

Fourier transforms of the power spectra give their respective correlation functions for the two- and three-dimensional fields: 
\begin{align}
 \xi^{\rm 2D}_{\rm XY}(r) &= \frac{1}{2 \pi} \int \! \mathrm{d} k \, k \, J_0(kr) \, P^{\rm 2D}_{\rm XY}(k), \nonumber\\
 \xi^{\rm 3D}_{\rm XY}(r) &= \frac{1}{2 \pi^2} \int \! \mathrm{d} k \, k^2 \, \frac{\sin(kr)}{kr} \, P_{\rm XY}(k), 
 \label{2d_fourier}
\end{align}
where $J_0$ is the zeroth-order Bessel function.
We note that the 2D and 3D power spectra follow a simple scaling relation, i.e., $P^{\rm 2D}_{\rm XY}(k)=P^{\rm 3D}_{\rm XY}(k)/L$ (where $L$ is the projection length), but the correlation functions do not obey this relation, $\xi^{\rm 2D}_{\rm XY}(r) \neq \xi^{\rm 3D}_{\rm XY}(r)/L$, from equation~(\ref{2d_fourier}).
Instead, $\xi^{\rm 2D}_{\rm XY} (r)$ simply scales as $\xi^{\rm 2D}_{\rm XY} (r) \propto L^{-1}$.
As we showed, the perturbation theory gives a validation of the response functions at least in the linear regime, so it is useful to compare with the simulation results. Performing the 2D Fourier transformation of the perturbation theory response of $P_{\rm XY}(k)$ (equation~(\ref{pk_resp_pert})) yields expressions for the response function of $\xi^{\rm 2D}_{\rm XY}(r)$ to $\delta_{\rm b}$: 
\begin{align}
 \frac{\mathrm{d} \xi{}_{\rm mm}^{\rm 2D}(r)}{\mathrm{d} \delta_{\rm b}} &= \frac{61}{21} \xi_{\rm m,lin}^{\rm 2D}(r) + \frac{1}{3} \frac{\mathrm{d} \xi_{\rm m,lin}^{\rm 2D}(r)}{\mathrm{d} \ln r},  \nonumber \\
 \frac{\mathrm{d} \xi_{\rm hm}^{\rm 2D}(r)}{\mathrm{d} \delta_{\rm b}} &= \left( \frac{61}{21} + \frac{b_2}{b_1} - b_1 \right) b_1 \xi_{\rm m,lin}^{\rm 2D}(r) + \frac{b_1}{3} \frac{\mathrm{d} \xi_{\rm m,lin}^{\rm 2D}(r)}{\mathrm{d} \ln r}, 
\label{xi2d_resp_pert}
\\
 \frac{\mathrm{d} \xi_{\rm hh}^{\rm 2D}(r)}{\mathrm{d} \delta_{\rm b}} &= \left( \frac{61}{21} + 2 \frac{b_2}{b_1} - 2 b_1 \right) b_1^2 \xi_{\rm m,lin}^{\rm 2D}(r) + \frac{b_1^2}{3} \frac{\mathrm{d} \xi_{\rm m,lin}^{\rm 2D}(r)}{\mathrm{d} \ln r}, \nonumber 
\end{align}
where $\xi^{\rm 2D}_{\rm m,lin}(r)$ is the linear projected correlation function of matter (obtained by plugging the linear matter power spectrum 
$P_{\rm m,lin}(k)$ into equation~(\ref{2d_fourier})). 
The above equation is similar to the response function for the 3D {$\xi(r)$} from \citet{2012PhRvD..85j3523S} \citep[see also][]{Baldaufetal:16}, except for numeric coefficients in the first term, i.e. $68/21$ for the 3D {$\xi(r)$} instead of $61/21$. 

Fig.~\ref{fig:xi2d_response} displays the $\xi^{\rm 2D}_{\rm XY}(r)$ responses obtained by performing 2D Fourier transformations of the $P^{\rm 2D}_{\rm XY}(k)$ response functions in Fig.~\ref{fig:pk2d_response}.
{The simulation results (denoted by the symbols) are obtained by performing the Fourier transform of measured power spectra from the simulations.}
The solid-red and dashed-orange curves are the perturbation theory predictions (equation~(\ref{xi2d_resp_pert})), which differ in the minimum wavenumber $k_{\rm min}$ in the integration when computing the Fourier transform. 
For the solid-red curves, we employed $k_{\rm min}=2\pi/L$, where $L=250 \, h^{-1}{\rm Mpc}$ is the box size of SU simulations, while we set $k_{\rm min}=0$ for the dashed-orange curves. 
Here we take into account a smoothing of the theoretical prediction due to the $r$-binning ($\Delta \log r =0.2$ in the figure). 
Clearly, the solid curves give a better match to the simulation results at large separations.  
Since the response function of halo-halo {$P(k)$} is noisy, owing to shot noise contamination, we did not find a reliable $k$-integration results; thus, it is not shown here. 
This agreement is not possible if we employ $68/21$ instead of $61/21$ for the coefficient of the first term in the perturbation theory prediction. 
This is also confirmed by Fig.~\ref{fig:xi3d_response}, which shows responses for {$\xi^{\rm 3D}_{\rm XY}(r)$}. 
Here, we performed 3D Fourier transforms of the 3D $P_{\rm XY}(k)$ response functions from Fig. \ref{fig:pk2d_response}. 
Perturbation theory predictions, with the coefficient $68/21$, agree well with SU simulation results at large separations.
{For $\xi(r)$ response, the simulation results are larger than the perturbation theory at small scales but, for $P(k)$ response, the simulations are smaller in Figure \ref{fig:pk2d_response}. This is because Figure \ref{fig:pk2d_response} is plotted in the fractional form, i.e., the response is divided by the nonlinear (linear) $P(k)$ for the simulations (perturbation theory).} 

\subsection{SSC calibration for cosmic shear correlation functions}
\label{sec:cosmic_shear}

\begin{figure}
    \begin{center}
    \vspace*{0cm}
	\includegraphics[width=0.5\columnwidth]{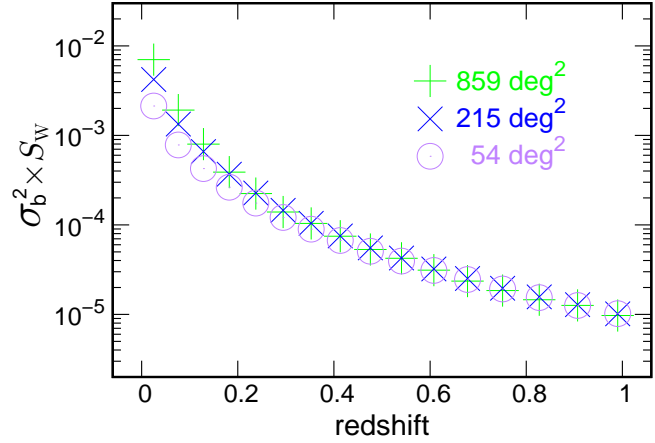}
    \caption{Variance of the large-scale density contrast in the shell-volume 
    along the line-of-sight direction, $\sigma_{\rm b}^2(z_i)$, for a fixed survey area, computed from the light-cone simulation {realisations} (see text for details).  
    Here, the shell-volume element at $z_i$ is given by $\Delta V(z_i)=S_W\chi_i^2\Delta \chi$, where $\chi_i$ is the radial distance to the $i$-th redshift and $\Delta \chi=150 \, h^{-1}{\rm Mpc}$ is the radial thickness fixed for all shells. 
    The different symbols show the results for $S_W\simeq 54, 215$ and 859 sq. degrees, respectively. We plot $S_W\times \sigma_{\rm b}^2(z_i)$ so that the results for different areas have similar amplitudes. There are $16$ lens shells from $z=0$ to $1$.
    }
    \label{fig:sigmab2}
    \end{center}
\end{figure}
\begin{figure}
    \vspace*{-4cm}
	\includegraphics[width=\columnwidth]{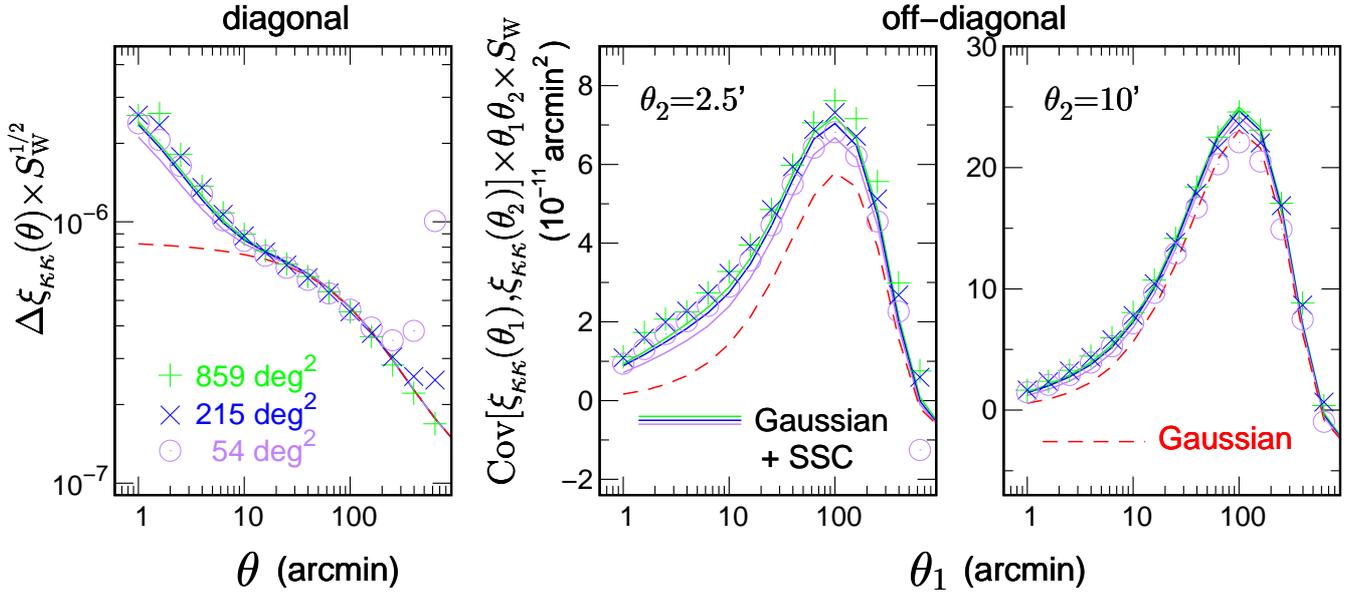}
	\vspace*{-2em}
    \caption{
    {\it Left panel}: The square root of the variance of cosmic shear two-point correlation function, i.e., the diagonal components of the covariance matrix. Here, source galaxies are at a single redshift, $z_{\rm s}=1.033$. The plus, cross and circle symbols are the results estimated from numerous realisations of a {mock} cosmic shear survey, which were constructed from full-sky ray-tracing simulations (see text for details). 
    The different symbols vary in the assumed survey areas as indicated, and 
    the survey geometry for each realisation is defined based on the \texttt{Healpix} subdivisions of the sky (see text for details). Simulation results include SSC contributions. We plot the results for $\Delta \xi^{\rm 2D}_{\kappa\kappa}\times S_W^{1/2}$, where $\Delta \xi^{\rm 2D}_{\kappa\kappa}\equiv {\rm Cov}^{1/2}$ and $S_W$ is the survey area, such that all results appear to have similar amplitudes. 
    The dashed red curves show the Gaussian error predictions (equation~(\ref{xi2d_cov_gauss})). 
    The solid curves show results obtained by adding the SSC contribution to the dashed curves for each survey area, where we used $\sigma_{\rm b}^2(z_i)$ from Fig.~\ref{fig:sigmab2} and equation~(\ref{xi2d_cov_ssc}). 
    The SSC results, even though inexpensively computed, exhibit excellent agreement with the simulation results. {\it Middle and right panels}: Similar results but for the off-diagonal components of the covariance matrices, 
    ${\rm Cov}[\xi^{\rm 2D}_{\kappa\kappa}(\theta_1),\xi^{\rm 2D}_{\kappa\kappa}(\theta_2)]$, for a given $\theta_2$ (as indicated) with varying $\theta_1$ in the $x$-axis. 
    We plot the results multiplied by $\theta_1\theta_2\times S_W$ for illustrative purposes. 
    The SSC calibration method displays excellent agreement with the simulation results. 
}
    \label{fig:xi_covariance}
\end{figure}

In this section, we consider an application of our method to calibrate the covariance matrix of cosmic shear correlation function. Cosmic shear or cosmological weak lensing is the distortion effect on the shapes of distant galaxies, caused by intervening mass distribution in large-scale structures.
The lensing {\it convergence} field characterises the cosmic shear effect, i.e., the weighted mass distribution integrated along the line-of-sight direction \citep[e.g.,][]{Bartelmann:2001}. 
The convergence field in an angular direction ${\bm \theta}$ is expressed based on the form of equation~(\ref{eq:2Ddeltam}):
\beq 
 \kappa({\bm \theta};z_{\rm s}) = \int_0^{\chi_{\rm s}} \! \mathrm{d}\chi \, f_\kappa(\chi)  \, \delta_{\rm m}(\chi {\bm \theta},\chi),
\eeq 
by choosing:
\beq 
  f_\kappa(\chi) = \frac{3 H_0^2 \Omega_{\rm m}}{2} a^{-1}\!(\chi)\chi \left(1-\frac{\chi}{\chi_{\rm s}}\right),
\eeq 
for $f_{\rm m}(\chi)$ (we use the notational convention, $f_\kappa$, instead of $f_{\rm m}$). Here, $\chi_{\rm s}$ is the distance to source galaxies, and we consider a single source redshift ($z_{\rm s}=1.033$) for simplicity. 
It is straightforward to include the source redshift distribution in the following discussion. 
Using Limber's approximation, the {\it angular} power spectrum of the cosmic shear field is expressed as
\beq 
  C_\ell^{\kappa \kappa}\!(z_{\rm s}) = \int_0^{\chi_{\rm s}} \! \mathrm{d} \chi \frac{1}{\chi^2} f_\kappa^2(\chi) \, P_{\rm mm} \!\left( k=\frac{\ell}{\chi},z \right).
\eeq 
where $\ell$ is an angular multipole. Since the angular correlation function of the cosmic shear field is given by the 2D Fourier transform of $C_\ell^{\kappa \kappa}$ under the flat-sky approximation, we can express its form via a discrete summation of thin-shell integration, similarly to equation~(\ref{eq:P2d_discrete}):
\begin{align}
 \xi^{\rm 2D}_{\kappa\kappa}\!(\theta; z_{\rm s}) &=  \int \! \frac{\ell \mathrm{d} \ell}{2\pi} \, C_\ell^{\kappa \kappa} (z_{\rm s}) \, J_0({\ell} \theta), \nonumber \\
 &= \int \! \frac{k\mathrm{d} k}{2\pi} \int \! \mathrm{d} \chi\,  f_\kappa^2(\chi) \, P_{\rm mm}(k,z)\, J_0(k \chi \theta),  \nonumber\\
 &\simeq \sum_{i=1}^{N_{\rm s}} \left[ f_\kappa(\chi_i) \, \Delta \chi \right]^2  \int \! \frac{k\mathrm{d}k}{2\pi} P^{\rm 2D}_{\rm mm}\!(k,z_i)\, J_0 (k \chi_i \theta), \nonumber \\
 &= \sum_{i=1}^{N_{\rm s}} \left[ f_\kappa(\chi_i;z_{\rm s}) \, \Delta \chi \right]^2 \xi^{\rm 2D}_{\rm mm}\!(\chi_i \theta,z_i), 
 \label{2d_conv_corr_func}
\end{align}
in which we use the relation $P^{\rm 2D}_{\rm mm}(k; z_i)=P_{\rm mm}(k; z_i)/\Delta\! \chi$ that is valid for a sufficiently thin shell around $\chi_i$ (or such a narrow shell width of $\Delta\!\chi$ should be used), and we assume that the radial thickness, $\Delta\!\chi$, is the same for all shells.
Summation continues to the $N_{\rm s}$-th shell corresponding to the source redshift.
For consistency with the notation used thus far, we use the superscript ``2D'' to denote the {\it projected} correlation function of cosmic shear, $\xi_{\kappa\kappa}^{\rm 2D}$. 
Therefore, the SSC term in the $\xi_{\kappa\kappa}^{\rm 2D}$ covariance is 
\beq 
 {\rm Cov}^{\rm SSC}\! \left[ \xi^{\rm 2D}_{\kappa\kappa}\!(\theta_1; z_{\rm s}), \xi^{\rm 2D}_{\kappa\kappa}\!(\theta_2; z_{\rm s}) \right] = \sum_i \left[ f_\kappa(\chi_i) \Delta \chi \right]^4 \frac{\mathrm{d} \xi_{\rm mm}^{\rm 2D}(\chi_i \theta_1,z_i)}{\mathrm{d} \delta_{\rm b}} \frac{\mathrm{d} \xi_{\rm mm}^{\rm 2D}(\chi_i \theta_2,z_i)}{\mathrm{d} \delta_{\rm b}} \sigma_{\rm b}^2(z_i).
\label{xi2d_cov_ssc}
\eeq 
Note that only the large-scale density variance, $\sigma_{\rm b}^2(z_i)$, depends on the survey-window function. 
To compute the above equation numerically, we first measured the responses from SU simulations at each of the $16$ redshifts\footnote{$z=0,0.05,0.10,0.15,0.20,0.25,0.31,0.36,0.42,0.48,0.55,0.62,0.69,0.77,0.85,$ and $0.93$.} and subsequently interpolated them to estimate the response at an arbitrary redshift. Redshift evolution and $k$-dependence of the response functions are all smooth, thus the interpolation is efficient. As given in Table~\ref{table1}, we use {\it Planck} cosmology for the SU simulations, while the light-cone ray-tracing simulations from \citet{Takahashietal:17}, which we use for comparison, are based on the nine-year WMAP cosmology \citep{Hinshaw:2013} (hereafter, {\it WMAP}). 
The correlation function for the {\it Planck} model is $12\%-23\%$ higher in amplitude than that for the {\it WMAP} model at $\theta=1'-100'$.
To correct for differences due to the different cosmological models, we multiply the {\it Planck}-based SSC prediction (equation~(\ref{xi2d_cov_ssc})) by a factor of $[\xi^{\rm 2D}_{\kappa\kappa}\!(\theta_1;z_{\rm s}) \xi^{\rm 2D}_{\kappa\kappa}\!(\theta_2;z_{\rm s})]_{\rm \it WMAP}/[\xi^{\rm 2D}_{\kappa\kappa}\!(\theta_1;z_{\rm s}) \xi^{\rm 2D}_{\kappa\kappa}\!(\theta_2;z_{\rm s})]_{\rm \it Planck}$.

To compute the total covariance matrix, we include the {G} {and SSC terms} but ignore the {cNG} term for simplicity. 
{One reason is the cNG contribution is an order of magnitude smaller than SSC from the left panels of Figure \ref{fig:pk2d_variance} (corresponding to cosmic shear variance), which show the errors in Case (iii) (= G+cNG) are much smaller than those in other cases (= G+cNG+SSC) in the non-linear regime.}
{Another} reason is we do not have sufficient simulation realisations to compute the {cNG} part (or equivalently, contributions from the connected parts of the four-point functions). We use an analytical formula to compute the {G} covariance 
\citep{Joachimi2008}:
\beq 
 {\rm Cov}^{\rm G}\!\left[ \xi^{\rm 2D}_{\kappa\kappa}\!(\theta_1; z_{\rm s}), \xi^{\rm 2D}_{\kappa\kappa}\!(\theta_2,z_{\rm s}) \right]
  = \frac{1}{ S_{\rm W}} \int \! \frac{\ell \mathrm{d} \ell}{\pi} \,  J_0(\ell \theta_1) \, J_0(\ell \theta_2) \left[ C_\ell^{\kappa\kappa}(z_{\rm s}) \right]^2\, ,
\label{xi2d_cov_gauss}
\eeq 
where $S_W$ is the survey area in units of steradians.
To compute this equation, we first employ the revised \texttt{halofit} \citep{Smith2003,Takahashi2012} to compute the nonlinear matter power spectrum for the {\it WMAP} cosmology and subsequently perform the $k$- and $\chi$-integrations to obtain the {G} error prediction. 
The {G} covariance scales with the survey area as $1/S_{W}$.

To test this SSC calibration method (equation~(\ref{xi2d_cov_ssc})), we use full-sky ray-tracing simulations developed in \cite{Takahashietal:17} \footnote{The full-sky light-cone simulation data (lensing fields and halo catalogues) are freely available for download at \url{http://cosmo.phys.hirosaki-u.ac.jp/takahasi/allsky_raytracing/}.}
to estimate the full covariance matrix. 
The full-sky simulation map (lensing fields) is given in the \texttt{Healpix} pixelisation \citep{Gorski2005}, in which the sphere is divided by $12 \times N_{\rm side}^2$ equal-area pixels.
We use maps with $N_{\rm side}=4096$ and $8192$, corresponding to a pixel size of $0.43 \, (N_{\rm side}/8192)^{-1}$ arcmin. 
In this study, we use $108$ full-sky lensing maps for source galaxies at a single redshift, $z_{\rm s}=1.033$.
{Each full-sky map contains $16$ spherical-lens planes placed at equal radial intervals of $\Delta \chi=150 \, h^{-1}{\rm Mpc}$ up to $z_{\rm s}=1.033$\footnote{The original full-sky maps that are publicly available have the lens planes up to the last scattering surface  ($z_{\rm s}=1100$).}. 
The projected density field on each lens plane is computed from the radial projection of matter distribution within the shell thickness.}
For each realisation, the halo catalogue containing information on mass and radial and angular positions for each halo is available\footnote{These halos are identified by the halo finder \texttt{Rockstar} \citep{Behroozi:2013} from the $N$-body simulations.}, and we use these lensing maps and halo catalogues to construct {mock} measurements of galaxy--galaxy lensing (shear-halo lensing) for the source galaxy sample \citep{Shirasakietal:17,2018arXiv180209696S}. 
To increase the number of realisations for {mock} measurements as well as to study the survey area dependence, we divide each full-sky map into $48,192$ and $768$ equal-area subregions according to the \texttt{Healpix} pixelisation, corresponding to 859, 215, and 54~square degrees for each subregion area, respectively. 
By considering each subregion to be one survey region, we have $5184,20736,$ and $82944$ samples from the $108$ full-sky map realisations in total, which is sufficient to precisely estimate the covariance matrix of cosmic shear correlation functions {$\xi_{\kappa\kappa}^{\rm 2D}$}. 
Since {$\xi_{\kappa\kappa}^{\rm 2D}$} are affected by super-survey modes, which exist across each subregion, the covariance matrix estimated in this manner should include all the contributions of {G}, {cNG} and SSC terms. 
When estimating {$\xi_{\kappa\kappa}^{\rm 2D}$} from a given survey region, we use the following estimator: 
\beq 
 \hat{\xi}^{\rm 2D}_{\kappa \kappa}(\theta; z_{\rm s}) = \frac{1}{N_\theta} \sum_{|{\bm \theta}_1-{\bm \theta_2}| \in \theta} \kappa({\bm \theta}_1; z_{\rm s}) \kappa({\bm \theta}_2; z_{\rm s}),
\eeq
where $\kappa({\bm \theta}_1;z_{\rm s})$ is the convergence field at ${\bm \theta}_1$, and $N_\theta$ is the number of pairs within $\theta-\Delta \theta/2 < |{\bm \theta}_1-{\bm \theta_2}| < \theta+\Delta \theta/2$ with a bin width of $\Delta \theta$. In the following, we ignore the shape noise contribution. 
\cite{Takahashietal:17} verified that the full-sky simulations provide accurate estimations of {$\xi_{\kappa\kappa}^{\rm 2D}$} and agree with analytical model predictions to within 5\% at $\theta>1'$ (see Section~3.2 of their paper).

Another model ingredient that is necessary in the SSC calibration method is the variance of super-survey density contrast at each {lens} shell along the line-of-sight direction, $\sigma_{\rm b}^2(z_i)$, in equation~(\ref{xi2d_cov_ssc}). 
To evaluate this variance, we need to properly use the survey-window function (angular and radial windows) at each redshift, $z_i$. Provided that a survey area is sufficiently wide, $\sigma_{\rm b}$ arises from the linear mass fluctuations and we can use the linear mass power spectrum to compute $\sigma_{\rm b}^2$ in equation~(\ref{eq:sigmab_W})
\citep[also see][]{TakadaHu:13,Lietal:14}. To minimize any possible uncertainty in the comparisons, we use the same subregions in the full-sky simulations to estimate $\sigma_{\rm b}^2(z_i)$ from the variance of the surface mass density on each shell, for a given survey area.
Fig.~\ref{fig:sigmab2} displays $\sigma_{\rm b}^2(z_i) \times S_W$ as a function of redshift.
Note that the redshift binning is fixed at $\Delta \chi=150 \, h^{-1}{\rm Mpc}$ as used in the full-sky ray-tracing simulations.
After multiplying $S_W$, all of the results for the different survey areas appear to be similar, although results do not exactly agree with each other, especially at low redshifts, because survey area dependence of $\sigma_{\rm b}$ originates from the shape of the linear {matter $P(k)$} \citep{TakadaHu:13,Takahashietal:14}.
We use the results of Fig.~\ref{fig:sigmab2} in equation~(\ref{xi2d_cov_ssc}) to estimate SSC contributions. 

Now we compare the covariance matrix estimated from the light-cone simulation realisations with that computed from the SSC calibration method.
Fig.~\ref{fig:xi_covariance}, in the left panel, displays the square-of-root of the variance, i.e., the diagonal covariance components. 
The different symbols represent the different survey areas, and we plot the results of $({\rm Cov}\times S_W)^{1/2}$ such that all results have similar amplitudes.  
The plot clearly shows that {NG} error is dominant at small scales, $\theta < 10'$, which is well explained by the SSC term in this method. While a close look reveals that the green curve is larger than the blue and purple curves, this is expected to be due to $\sigma_{\rm b}^2(z)$ at $z \lesssim 0.2$, as shown in Fig.~\ref{fig:sigmab2}.
At large scales, $\theta > 10'$, simulation results are consistent with the {G} prediction.
Deviation of the simulation results from the {G} error at large separations, particularly for small area cases, is due to boundary effects in the finite survey area that are missing in the analytical calculation in equation~(\ref{xi2d_cov_gauss}) \citep[see Appendix~A in][]{Satoetal:11}.
The middle and right panels display the off-diagonal components of the covariance matrix for a chosen $\theta_2$, but with varying $\theta_1$ in the $x$-axis. 
For $\theta_2=2.5'$, {NG} errors appear dominant but can similarly be explained by addition of the SSC term.
Note that these results vary with the source redshift or, more generally, the distribution of source redshifts, but it is straightforward to include these effects in the SSC calibration method.

\subsection{SSC calibration for halo-convergence correlation functions}
\label{sec:gg_lens}

\begin{figure}
    \vspace*{-4cm}
	\includegraphics[width=\columnwidth]{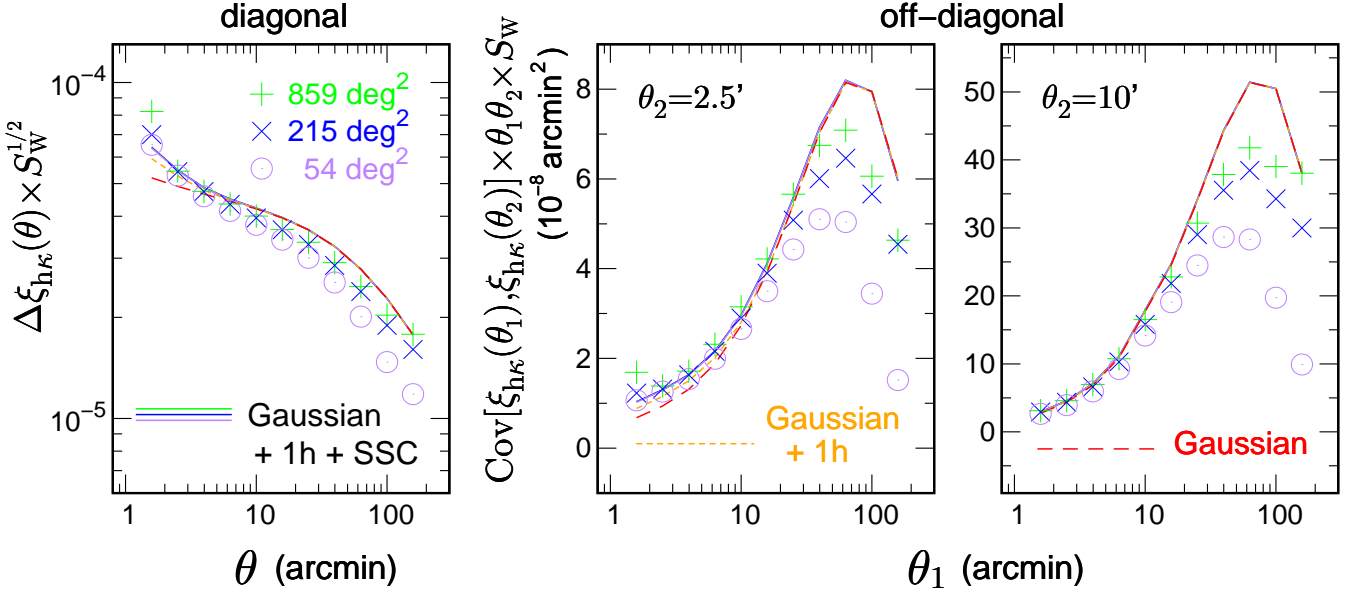}
	\vspace*{-2em}
    \caption{Similar to Fig.~\ref{fig:xi_covariance}, but this figure shows results for the covariance of the halo-convergence correlation function, which is relevant for galaxy--galaxy weak lensing, for a halo sample with $M>10^{13}h^{-1} {\rm M}_\odot$ in the redshift range of $0.47<z<0.59$. 
We employed $z_{\rm s}=1.033$ for the source redshift. Note that, in this study, we do not consider the shape {noise} contribution. 
The dashed red curves are the analytical Gaussian error prediction, the dotted orange curves include the one-halo term of trispectrum for the connected non-Gaussian covariance, and the solid curves further include the SSC contribution. 
On the $x$-axis, the angular separation, $\theta$, corresponds to the transverse comoving distance, $\simeq 0.4 \, h^{-1} {\rm Mpc} \, (\theta/1')$.
}
    \label{fig:gglens_covariance}
\end{figure}

In this section, we discuss the covariance for the angular correlation functions of foreground halos and the background convergence fields that is closely related to that measured in galaxy--galaxy lensing. 
We consider a sample of halos selected by mass above a given threshold, $M>M_{\rm th}=10^{13} h^{-1} {\rm M}_\odot$, in a redshift range of $0.47<z<0.59$, which mimics halos hosting BOSS CMASS galaxies \citep[e.g.][]{Alam:2015,Moreetal:2015}.
The number of halos is approximately $2.2 \times 10^6$ in the {full sky}, and the comoving number density $\bar{n}_{\rm h}\simeq 3.2\times 10^{-4}~h^3 {\rm Mpc}^{-3}$, which is similar to the number density of CMASS galaxy {sample ($\sim 3 \times 10^{-4}~h^3 {\rm Mpc}^{-3}$) in \cite{Miyatake:2015}}. 

The halo-convergence angular correlation function for the mass limited sample is defined by equations~(\ref{eq:2Ddeltah}) and (\ref{eq:2Ddeltam}) as follows:
\begin{align}
 \xi^{\rm 2D}_{{\rm h}\kappa}(\theta; z_{\rm l},M_{\rm th}) 
 &= \left. \langle \delta_{\rm h}^{\rm 2D}\!({\bm \theta}_1; z_{\rm l}, M_{\rm th}) \kappa({\bm \theta}_2;z_{\rm s}) \rangle \right|_{|{\bm \theta}_1-{\bm \theta}_2|=\theta},
 \nonumber \\
 &= \frac{1}{\bar{n}_{\rm h}^{\rm 2D}} \int \! \mathrm{d} \chi_1 \, f_{\rm h}(\chi_1) \int \! \mathrm{d} \chi_2 \, f_\kappa(\chi_2) \left. \int_{M_{\rm th}}^\infty \! \mathrm{d} M \frac{\mathrm{d} n}{\mathrm{d} M} (M,\chi_1) \langle \delta_{\rm h}(\chi_1 {\bm \theta}_1,\chi_1;M) \, \delta_{\rm m}(\chi_2 {\bm \theta}_2,\chi_2) \rangle \right|_{|{\bm \theta}_1-{\bm \theta}_2|=\theta}, \nonumber \\
 &\simeq \frac{1}{\bar{n}_{\rm h}^{\rm 2D}} \int \! \mathrm{d} \chi \, f_{\rm h}(\chi)  f_\kappa(\chi) \int_{M_{\rm th}}^\infty \! \mathrm{d}M\frac{\mathrm{d}n}{\mathrm{d}M}(M,\chi) \, \Delta \chi \int \!\frac{k\mathrm{d}k}{2\pi} P_{\rm hm}^{\rm 2D} (k;\chi) \, J_0(k\theta),
\nonumber\\
 &\simeq \frac{1}{\bar{n}_{\rm h}^{\rm 2D}} \sum_{i} \left[f_{\rm h}(\chi_i)f_{\kappa }(\chi_i) \, \bar{n}_{\rm h}(M_{\rm th};\chi_i) \, (\Delta \chi)^2\right] \xi_{\rm hm}^{\rm 2D} (\chi_i \theta;\chi_i,M_{\rm th}), 
\end{align}
where $f_{\rm h}(\chi)= \chi^2$ if $\chi$ is in the range of $0.47< z< 0.59$, otherwise $f_{\rm h}(\chi)=0$, $z_{\rm l}$ denotes the redshift range of lensing halos, and the projected halo number density, $\bar{n}^{\rm 2D}_{\rm h}$, given in equation~(\ref{halo_propaties}), has the dimensions of angular number density, i.e., $[{\rm rad}^{-2}]$. We also introduce the following quantities for notational simplicity:
\begin{align}
&\xi_{\rm hm}^{\rm 2D} (r;\chi, M_{\rm th})\equiv \frac{1}{\bar{n}_{\rm h}(M_{\rm th},\chi)} \int_{M_{\rm th}}^\infty \!\mathrm{d}M  \frac{\mathrm{d}n}{\mathrm{d}M}(M,\chi) \, \xi_{\rm hm}^{\rm 2D} (r;\chi, M), \nonumber\\
& \bar{n}_{\rm h}(M_{\rm th};\chi) \equiv \int_{M_{\rm th}}^\infty \!\mathrm{d}M  \frac{\mathrm{d}n}{\mathrm{d}M}(M,\chi) . 
\end{align}
where $\xi_{\rm hm}^{\rm 2D}(r;\chi, M_{\rm th})$ is the projected halo-matter correlation function for a mass limited sample, and $\bar{n}_{\rm h}(M_{\rm th};\chi)$ is the cumulative halo number density. 
Here we consider the convergence field for simplicity, but the following discussion is applicable to the lensing shear field by replacing $J_0(\ell \theta)$ in the equations with $J_2(\ell\theta)$ \citep[e.g.,][]{Hikage:2013}.

The SSC covariance term is derivable, similarly to equation~(\ref{xi2d_cov_ssc}), as
\beq
 {\rm Cov}^{\rm SSC}\! \left[ \xi^{\rm 2D}_{{\rm h}\kappa}\!(\theta_1; z_{\rm l}), \xi^{\rm 2D}_{{\rm h}\kappa}\!(\theta_2; z_{\rm l}) \right] = 
 \frac{1}{\left( \bar{n}^{\rm 2D}_{\rm h} \right)^2} \sum_i 
 \left[ f_{\rm h}(\chi_i) f_\kappa(\chi_i) \, \bar{n}_{\rm h}(M_{\rm th};\chi_i)
(\Delta \chi)^2 \right]^2 
\frac{\mathrm{d} \xi_{\rm hm}^{\rm 2D}(\chi_i \theta_1;z_i)}{\mathrm{d} \delta_{\rm b}} \frac{\mathrm{d} \xi_{\rm hm}^{\rm 2D}(\chi_i \theta_2;z_i)}{\mathrm{d} \delta_{\rm b}} \sigma_{\rm b}^2(z_i).
\label{eq:cov_ssc_xi2Dhm}
\eeq
Similarly to the previous case, we multiply the ratio 
$[\xi^{\rm 2D}_{{\rm h}\kappa}\!(\theta_1;z_{\rm l}) \, \xi^{\rm 2D}_{{\rm h}\kappa}\!(\theta_2;z_{\rm l})]_{\rm \it WMAP}/[\xi^{\rm 2D}_{{\rm h}\kappa}\!(\theta_1;z_{\rm l}) \, \xi^{\rm 2D}_{{\rm h}\kappa}\!(\theta_2;z_{\rm l})]_{\rm \it Planck}$ by the SSC above, to account for the differences in the cosmological model.

The {G} covariance was derived by \citet{JKJ:2009} \citep[also see][]{OguriTakada:2011} as
\beq
  {\rm Cov}^{\rm G}\!\left[ \xi^{\rm 2D}_{{\rm h} \kappa}(\theta_1;z_{\rm l}), \xi^{\rm 2D}_{{\rm h} \kappa}(\theta_2;z_{\rm l}) \right]
   = \frac{1}{ S_{\rm W}} \int \! \frac{\ell \mathrm{d} \ell}{\pi} \,  J_0(\ell \theta_1) \, J_0(\ell \theta_2) \left[ 
   \left\{C_\ell^{\rm hh}(z_{\rm l}) +\frac{1}{\bar{n}^{\rm 2D}_{\rm h}} \right\}\, C_\ell^{\kappa \kappa}(z_{\rm s}) + \left\{ C_\ell^{{\rm h} \kappa}(z_{\rm l}) \right\}^2 \right].
\eeq
To compute this contribution, we use the halo model \citep{2002PhR...372....1C} to evaluate the halo auto-power spectrum, $C^{\rm hh}_\ell$, and the halo-convergence cross-power spectrum, $C^{{\rm h}\kappa}_\ell$.
We adopted the same model parameters that were used in \cite{Oguri:2011}.
\cite{Takahashietal:17} verified that the halo model agreed with the full-sky simulation results for the halo auto-correlation function and the halo-convergence cross-correlation function (see Sections 3.3 and 3.4 of their paper). 

To obtain a better agreement between the SSC calibration method and the light-cone simulations, as we show below, we also include the {cNG} term, given in \cite{Satoetal:11} as follows:
\beq
 {\rm Cov}^{\rm cNG}\! \left[ \xi^{\rm 2D}_{{\rm h} \kappa}(\theta_1;z_{\rm l}), \xi^{\rm 2D}_{{\rm h} \kappa}(\theta_2;z_{\rm l}) \right] = 
 \frac{1}{4 \pi^2S_W} \int \! \ell_1 \mathrm{d} \ell_1 J_0(\ell_1 \theta_1) \int \! \ell_2 \mathrm{d} \ell_2 J_0(\ell_2 \theta_2) \, 
 \bar{T}_{{\rm h}\kappa{\rm h}\kappa}(\ell_1,\ell_2),
\eeq
where $\bar{T}_{{\rm h}\kappa{\rm h}\kappa}$ is the angle-averaged trispectrum of halo-convergence-halo-convergence. We again employ the halo model given in Appendix~A {of} \citet{KrauseEifler:16} to compute the trispectrum. In this study, we consider only the 1-halo term.

To test the SSC calibration method, we use light-cone weak lensing maps and halo catalogues constructed from the same $N$-body simulations \citep{Takahashietal:17}.
Here, we used $40$ full-sky maps with $N_{\rm side}=8192$ (corresponding to $\sim 0.43$~arcmin for the angular resolution).
We use the following halo-matter cross-correlation estimator: 
\beq
 \hat{\xi}^{\rm 2D}_{{\rm h}\kappa}(\theta;z_{\rm l}) = \frac{1}{N_{\rm h}} \sum_{i=1}^{N_{\rm h}} \frac{1}{N_{{\rm pair},i}} \!\!\!
 \left.\sum_{j=1}^{N_{{\rm pair},i}} \kappa({\bm \theta}_j;z_{\rm s})\right|_{|{\bm \theta}_{{\rm h},i}-{\bm \theta}_j| \in \theta}\, 
 {- \frac{1}{N_{\rm ran}} \sum_{i=1}^{N_{\rm ran}} \frac{1}{N^{\rm ran}_{{\rm pair},i}} \!\!\!
 \left.\sum_{j=1}^{N^{\rm ran}_{{\rm pair},i}} \kappa({\bm \theta}_j;z_{\rm s})\right|_{|{\bm \theta}_{{\rm ran},i}-{\bm \theta}_j| \in \theta}\,}
\label{xi_halokappa}
\eeq
where, {in the first term,} $N_{\rm h}$ is the number of halos in the survey region, the summation, $\sum_{j}$, runs over pairs between the $i$-th halo and the $j$-th pixel, satisfying the angular separation condition, $|{\bm \theta}_{{\rm h},i}-{\bm \theta_j}| \in \theta$, to within a bin width of  $\Delta \theta$, and $N_{{\rm pair},i}$ is the number of pairs for the $i$-th halo. 
Here, the halos are taken from the survey region, whereas the convergence fields are taken from the {full sky} (including outside the survey region).
As indicated by \citet{Singhetal:16} \citep[also see][]{Shirasakietal:17}, using a sufficient number of random catalogues is essential to have an unbiased estimate of {$\xi^{\rm 2D}_{{\rm h}\kappa}$} as well as the correct covariance matrix. We use $10$ times the number of random points than the number of halos in each realisation.
{The second term of equation (\ref{xi_halokappa}) represents this random point-convergence correlation: $N_{\rm ran}$, $N^{\rm ran}_{{\rm pair},i}$ and ${\bm \theta}_{{\rm ran},i}$ are the same quantities as in the first term but for the random points.} 

\begin{figure}
    \begin{center}
    \vspace*{0cm}
	\includegraphics[width=0.8\columnwidth]{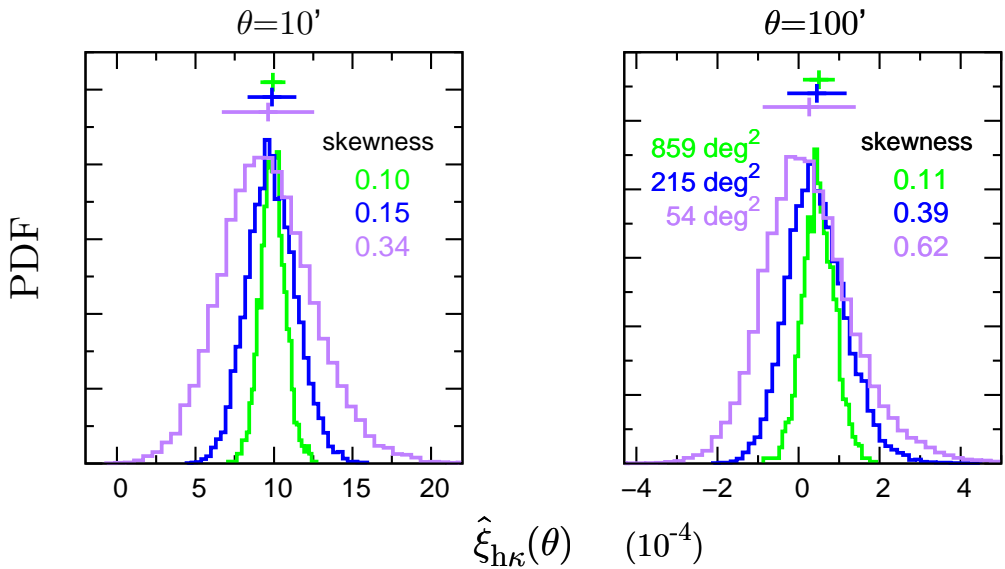}
	\vspace*{0cm}
    \caption{Probability distribution of $\hat{\xi}_{{\rm h}\kappa}(\theta)$ at $\theta=10'$ (left panel) and $100'$ (right) for the three survey areas. The PDF amplitudes are arbitrarily scaled. In each panel, we also show the mean with $1\sigma$ errors (the vertical lines with horizontal bars in the upper part) as well as the skewness, defined as $S_3=\langle (\hat{\xi}_{{\rm h}\kappa}- \langle \hat{\xi}_{{\rm h}\kappa} \rangle)^3 \rangle / \langle(\hat{\xi}_{{\rm h}\kappa}- \langle \hat{\xi}_{{\rm h}\kappa} \rangle)^2 \rangle^{3/2}$, which is exactly zero if the PDF follows the Gaussian distribution. For large $\theta$, the distribution is highly, positively skewed particularly for small survey areas. 
}
    \label{fig:gglens_pdf}
    \end{center}
\end{figure}

Fig.~\ref{fig:gglens_covariance} shows the SSC calibration method for the covariance of 
$\xi^{\rm 2D}_{{\rm h}\kappa}(\theta)$. First, we find that, compared to Fig.~\ref{fig:xi_covariance}, the {NG} covariance contribution is small and is significant only at small angular scales, $\theta\simlt 5'$.
The figure shows that the SSC calibration method fairly well  reproduces the covariance from the ray-tracing simulations at the small scales, if the {cNG} term is added. 
The one-halo term provides a moderate contribution, which is expected from the top-middle panel of Fig.\ref{fig:pk2d_variance}, showing the {cNG} error is significant at small scales (i.e. the circles are much larger than the horizontal dashed line).
But it is not the case in the left panels of the same figure (the matter-matter components, corresponding to cosmic shear). 
The analytical Gaussian covariance prediction appears to over-estimate covariance amplitudes at large scales, which are more clearly seen in the middle and right panels. This is owing to analytical calculations that do not consider survey geometry effects, which can be easily taken into account using similar methods found in \cite{Satoetal:11} 
\citep[see also][]{Shirasakietal:17,Murataetal:17,2018PASJ...70S..25M}. 
If there are a sufficient number of independent modes in a survey area, the distribution of $\hat{\xi}_{{\rm h}\kappa}$ approaches the Gaussian owing to the central limit theorem.
However, in our halo sample, there are less modes in the smaller survey areas at large scales, which cause the distribution to be highly skewed (see Fig. \ref{fig:gglens_pdf}). 
This figure clearly shows that skewness causes a smaller mean and variance especially for the smaller survey area.  
However the scope of this study is the {NG} covariance; therefore, this is not further explored. 

\section{Discussion and Conclusions}
\label{sec:conclusion}

In this study, we have studied the response functions for the projected power spectra {$P^{\rm 2D}(k)$} of matter and halos to large-scale density contrast ($\delta_{\rm b}$) and the covariance matrices for {$P^{\rm 2D}(k)$}. 
First, by using paired SU simulation realisations, we calibrated the response functions of {$P^{\rm 2D}(k)$} and subsequently compared the results with the response functions of their respective three-dimension power spectra as well as with perturbation theory predictions. 
We showed that the fractional response functions, $\mathrm{d}\ln P^{\rm 2D}_{\rm XY}(k)/\mathrm{d}\delta_{\rm b}$, are identical to their respective three-dimensional power spectra. 
In other words, the line-of-sight projection does not change the form of the response function, contrary to claims in analytical calculations based on Limber's approximation \citep{TakadaJain:09}. 
Simulation results match perturbation theory predictions at small $k$-bins, supporting this finding. 
Second, we studied the {$P^{\rm 2D}(k)$} covariance matrices (for matter-matter, matter-halo and halo-halo) using a sufficient number of realisations of projected fields constructed from $N$-body simulations. To do this, we used the following sets of projected field realisations: one set of realisations follow the periodic boundary conditions (i.e., no super-survey mode) and another set has super-survey mode contributions that are in both parallel and perpendicular directions to the line-of-sight (projection) direction, and the final set has only the super-survey modes in the perpendicular direction. 
{We showed that the SSC calibration method enables to reproduce $P^{\rm 2D}(k)$ covariance, in which the SSC contribution based on the response function is added to the covariance from simulations with periodic boundary conditions.}
This is analogous to what was shown in \citet{Lietal:14} for {3D} matter {$P(k)$}.
We did not observe a clear signature of the super-survey mode effect in the line-of-sight direction or the super-survey tidal effect, within the statistical accuracy of current simulation realisations {(about $10 \, \%$)}. 
{Although this does not seem} consistent with recent claims made by \cite{2017arXiv171107467B}, a further careful study is needed to resolve this inconsistency. 
Third, we used the response {function} method to calibrate the covariances for the cosmic shear correlation function and the halo-convergence cross-correlation function, {the latter of which is essential for galaxy--galaxy lensing analyses}. We showed that the response calibration method {accurately} reproduces the covariance obtained from the light-cone ray-tracing simulations. 

Thus, the response calibration method developed in this paper offers an efficient technique to calibrate covariance matrices for the projected field, without running light-cone ray-tracing simulations. To obtain a sufficiently accurate calibration of the covariance matrix for ongoing and upcoming wide-area galaxy surveys, the following can be a practically useful method:
\begin{itemize}
\item {\bf Gaussian covariance contribution} (${\rm Cov}^{\rm G}$) -- {This term consists of two parts: sample variance and shot noise.} This is the simplest and is naively considered to be the easiest to calibrate. 
However, calibration requires precise care because one needs properly take into account observational effects such as survey geometry/masks and the intrinsic distribution of galaxies used in the sample. To do this, we should use the real galaxy catalogue as much as possible, as several studies have already done \citep[][]{Shirasakietal:17,Murataetal:17,2018PASJ...70S..25M}. 
For example, for the lensing field, the real catalogue of source galaxies should be used as follows: (1) Randomly rotate galaxy ellipticity for each galaxy to erase real lensing effects.
(2) Apply actual lensing measurement pipeline to the modified galaxy catalogue to estimate correlation functions such as the cosmic shear correlation and galaxy--galaxy lensing. 
(3) Repeat the first and second procedures, and then estimate {the shot noise part of the G covariance from the rotated data.} 
The estimated covariance matrix includes all of the observational effects, such as survey masks, intrinsic shape distribution 
and photometric redshift distribution. 
In particular, this covariance can take into account the correlated shape noise, which arises because ellipticities of the same source galaxies, i.e., galaxies with relatively large intrinsic ellipticities, are used multiple times in the correlation function estimation.  
To further include the {G} sample variance, which arises from products of the power spectrum or two-point correlation function, we could use the method in Appendix~A of \citet{Satoetal:11}; they developed a method to estimate the {G} sample variance taking into account actual pairs used in the analysis. It would be straightforward to extend this method to including cross terms between the shape noise and the sample variance.
Or one could use the Gaussian realisations of lensing fields to estimate the {G} sample variance, and then add contribution to the shape noise covariance. This would be our future work. 

\item {\bf Connected non-Gaussian contribution} (${\rm Cov}^{\rm cNG}$) -- This term becomes non-negligible only over the transition regime (which is a narrow scale range) between the {G} covariance and the SSC contribution. 
In addition, survey {window} effects can be ignored in this contribution if separations are much smaller than the survey-window size. Hence, to calibrate this term, one could use a set of $N$-body simulations with periodic boundary conditions as we have done in this paper, or use the halo model predictions as done in \citet{Satoetal:09} \citep[see also][]{CoorayHu:01,TakadaHu:13,Lietal:14,KrauseEifler:16}.

\item {\bf Super-sample covariance contribution} (${\rm Cov}^{\rm SSC}$) -- SSC contribution can be significant at small scales. As we have shown in this study, we can use the response function approach to calibrate the SSC term. To obtain the response function for a given observable, we use SU simulations, which are quite efficient at reducing sample variance contamination. 
In principle, one could use only a paired simulation to calibrate the response function as a function of redshifts and separation scales for a given cosmological model. 
This is quite inexpensive for current computer resources. Then, one could compute the SSC contribution from the line-of-sight integration of the response function, weighted by the proper radial functions (e.g., see equations~(\ref{xi2d_cov_ssc}) and (\ref{eq:cov_ssc_xi2Dhm})). 
The SSC term also originates from the line-of-sight integral of the variance of density contrast averaged in the survey area, $\sigma_{\rm b}^2(z_i; S_W)$, at each redshift. Only this term depends on the survey window (survey masks) and can be computed from the integration of linear power spectrum or the linear Gaussian realisations \citep[see also][for the similar discussion]{TakadaHu:13,Lietal:14}.
Thus, this method does not require light-cone simulations. 
\end{itemize}
Thus, by combining the three contributions, one could calibrate the full covariance matrix for a given observable. This is a computationally inexpensive method, and it allows one to include the cosmological dependence of the covariance matrix in the parameter estimation. To do this, we need to calibrate the response function of a given observable as a function of cosmological parameters. Nishimichi et al. (in prep.) are now constructing an {\it emulator} to output the power spectra for matter and halos as functions of halo mass, redshift and cosmological model. They are planning to include the response functions in the emulator, which could allow one to calibrate the covariance matrices for matter and halo observables. However, we do not wish to claim that light-cone simulations are no longer needed \citep{Takahashietal:17}. Such simulations are still required to make mock catalogues of a galaxy survey, which are useful for testing systematic errors and analysis pipelines and for calibrating the covariance matrix. However, since making a large number of light-cone simulations for different cosmological models is still computationally expensive, these methods are complementary to each other. 

In this paper, we have ignored the effects of large-scale tidal fields on {$P^{\rm 2D}(k)$}. Although we did not find a clear signature of these effects, it is interesting to further explore whether the large-scale tidal fields affect {$P^{\rm 2D}(k)$} because this carries independent information about large-scale gravitational fields.
As pointed out by \citet{Akitsuetal:17} \citep[see also][]{Akitsu:2017b}, large-scale tidal fields directly affect the redshift-space {galaxy} power spectra, or, more generally, the redshift-space clustering features.
It would be straightforward to extend the method in this paper to develop a method of calibrating the covariance of the redshift-space clustering power spectra based on the response approach. However, in this case, the large-scale tidal field is a tensor, i.e. depends on directions and vary with the degree of alignment with the line-of-sight direction. When we need to consider a wide-area galaxy survey, effects of spherical curvature needs to be taken into account because the directions of the tidal tensor relative to the line-of-sight direction would vary with redshift. These would be interesting subjects and worth exploring.

\section*{Acknowledgements}
We thank Elisabeth Krause and Fabian Schmidt for their useful discussions. This work was in
part supported by Grant-in-Aid for Scientific Research from the JSPS
Promotion of Science (No.~23340061, 26610058, 15H03654, and 17H01131), MEXT
Grant-in-Aid for Scientific Research on Innovative Areas (No.~15H05887, 15H05893,
15K21733, and 15H05892), JSPS Program for Advancing Strategic
International Networks to Accelerate the Circulation of Talented Researchers,
and Japan Science and Technology Agency (JST) CREST Grant Number JPMJCR1414.
Numerical computations were in part carried out on Cray XC30 at Centre for Computational Astrophysics, National Astronomical Observatory of Japan.

\bibliographystyle{mnras}
\bibliography{refs} 




\appendix


\section{Separation Universe Method}

{
In this appendix, we present a short summary of the separate universe (SU) technique developed by, e.g., \cite{Sirko:2005}, \cite{Lietal:14}, and \cite{Baldaufetal:16}.
The SU method is frequently used to calculate the $P(k)$ response and the resulting SSC term.   
We consider a large over/under-density region with the mean density contrast, $\delta_{\rm b}$, with respect to the global mean (assuming $|\delta_{\rm b}| < 1$).
We set up a cubic simulation box with constant $\delta_{\rm b}$ and follows the gravitational evolution of dark matter particles by an $N-$body method.
This procedure is the same as the standard $N-$body simulation but the background mean density simply shifts by constant value $\delta_{\rm b}$. 
We choose three $\delta_{\rm b}$ values at present ($z=0$): $\delta_{\rm b}=\pm 0.01$ (the paired SU simulations) and $0$ (the default setting).
The box sizes are chosen to be identical in all the cases, $L=L_W=250h^{-1} \, {\rm Mpc}$.
Here, a quantity with (or without) a subscript ``W'' means a value in SU (or the global background universe).
The cosmological parameters ($\Omega_{{\rm m}W}, \Omega_{\Lambda W}, \Omega_{{\rm K}W}, h_W$) and the scale factor, $a_W$, are different from those in the global background, but the time ($t$) is the same in both frames.
We will derive relations between those parameters with and without ``W'' below.
The mean matter density in the SU simulation box, $\bar{\rho}_{{\rm m}W}$, is related to the global mean, $\bar{\rho}_{\rm m}$, via 
\beq
 \bar{\rho}_{{\rm m}W} (a_W)  = \bar{\rho}_{\rm m} (a) \left[ 1+\delta_{\rm b}(a) \right].
 \label{eq:Li30}
\eeq
The density contrast, $\delta_{\rm b}(a)$, grows proportional to the linear growth factor, $D(a)$, in the global cosmology as long as it stays in the linear regime:
\beq
  \delta_{\rm b}(a)= \frac{D(a)}{D_0} \, \delta_{{\rm b}0},  ~~  D(a) = \frac{5 \Omega_{\rm m}}{2} \frac{H(a)}{H_0} \int_0^a \!\! da^\prime \frac{H_0^3}{a^{\prime 3} H^3(a^\prime)},
  \label{eq:lgf}
\eeq
where the subscript $0$ means the present values and $D$ is normalized as $D \rightarrow a$ at $a \rightarrow 0$.
In our setting of $\delta_{{\rm b}0}=\pm 0.01$, the above linear approximation is valid.
Our $N$-body code (based on \texttt{Gadget2}) solves the non-linear evolution of $\delta_{\rm b}$ using the spherical collapse model but the non-linear correction is very small.
The present matter density parameter, $\Omega_{\rm m}$, is defined in a standard manner as,
\beq
 \Omega_{\rm m} = \frac{8 \pi}{3 H_0^2} \, \bar{\rho}_{{\rm m}0} = \frac{8 \pi}{3 H_0^2} \, \bar{\rho}_{\rm m}(a) \, a^3.
 \label{eq:Omegam}
\eeq
Therefore we have $\bar{\rho}_{\rm m}(a) \propto \Omega_{\rm m}h^2 a^{-3}$, and the same relation holds in the SU frame: $\bar{\rho}_{{\rm m}W}(a_W) \propto \Omega_{{\rm m}W}h_W^2 a_W^{-3}$.
From these relations, equation (\ref{eq:Li30}) reduces to
\beq
 \frac{\Omega_{{\rm m}W} h_W^2}{a_W^3} = \frac{\Omega_{\rm m} h^2}{a^3} \left[ 1+\delta_{\rm b}(a) \right].
 \label{eq:Li31}
\eeq
In the limit of high redshift ($a \rightarrow 0$), the scale factors should be identical ($a_W \rightarrow a$) and $\delta_b \rightarrow 0$ from equation (\ref{eq:lgf}).
Therefore, we have $\Omega_{{\rm m}W} h_W^2=\Omega_{\rm m} h^2$ from equation (\ref{eq:Li31}), and inserting this into equation (\ref{eq:Li31}) we have a relation between  the scale factors as
\beq
 a_W = \frac{a}{\left[ 1+\delta_{\rm b}(a) \right]^{1/3}} \simeq \left[ 1 - \frac{1}{3} \delta_{\rm b}(a) \right] a.
 \label{eq:Li35}
\eeq
We used the above equation to calculate the output redshifts in the SU simulations from those in the global background ($z=0.20$ and $0.55$ in our case). 
}

{We next consider a relation of comoving lengths $r_W$ and $r$.
As these physical lengths are identical, we have $a_W r_W = a r$ or
\beq
 r_W = \frac{a}{a_W} r \simeq \left[ 1 + \frac{1}{3} \delta_{\rm b}(a) \right] r.
 \label{eq:comv_su}
\eeq
A similar relation holds for comoving wavevectors as $a_W k_W^{-1} = a k^{-1}$ or
\beq
 k_W = \frac{a_W}{a} k \simeq \left[ 1 - \frac{1}{3} \delta_{\rm b}(a) \right] k.
 \label{eq:Li44}
\eeq
The above equation (\ref{eq:Li44}) should be used when converting $k_W$ measured in the SU simulations to $k$ in the global frame.
We then discuss a relation between power spectra $P_{{\rm XY}W}(k_W)$  and $P_{\rm XY}(k)$ for matter (X,Y=m) or halo (X,Y=h).
Since the comoving scales are different from equation (\ref{eq:comv_su}) and the power spectra have dimensions of (comoving length)$^3$, we employ the dimensionless power spectra, $\Delta_{{\rm XY}W}^2(k) \, (= k_W^3 P_{{\rm XY}W}(k_W)/(2 \pi^2))$ and $\Delta_{\rm XY}^2(k)$, to erase the difference, following \cite{Lietal:14}. 
Furthermore, the matter density fluctuations in each frame are different by a factor of $(1+\delta_{\rm b})$ due to the different background density (see also main text after equation (\ref{pkresp_relation})). 
Therefore, these power spectra are related via,
\beq
  \Delta_{\rm XY}^2(k) = \left( 1+\delta_{\rm b} \right)^n \Delta_{{\rm XY}W}^2(k_W),
\eeq
which is the same as equation (\ref{pkresp_relation}).
In summary, we measured the power spectra $P_{{\rm XY}W}(k_W)$ in the paired SU simulation with $\delta_{{\rm b}0}=\pm 0.01$, and calculated the numerical derivative to $\delta_{\rm b}$ using equation (\ref{eq:3D_response}) to obtain the $P(k)$ response. 
}

{Finally, we move on to the other cosmological parameters. The Hubble parameters in both frames are defined as, $H=\dot{a}/a$ and $H_W=\dot{a}_W/a_W$, and the latter is given from equation (\ref{eq:Li35}) as,
\beq
 H_W(a_W) = \frac{\dot{a}_W}{a_W} = \frac{\dot{a}}{a} - \frac{1}{3} \frac{\dot{\delta}_{\rm b}(a)}{1+\delta_{\rm b}(a)} \simeq H(a) - \frac{1}{3} \dot{D}(a) \, \delta_{{\rm b}0}.
 \label{eq:Li36}
\eeq
Using the Hubble equation, $H^2(a) = H_{0}^2 [ \Omega_{\rm m} a^{-3} + \Omega_{\rm K} a^{-2} + \Omega_\Lambda ]$, and equations (\ref{eq:lgf}) and (\ref{eq:Li36}), we have 
\beq
 H_W^2(a_W) \simeq H^2(a) + \frac{H_{W0}^2 - H_0^2}{a^2} + H_0^2 \delta_{\rm b}(a) \left( \frac{\Omega_{\rm m}}{a^3} + \frac{2}{3} \frac{\Omega_{\rm K}}{a^2} \right).
 \label{eq:Li39}
\eeq
The above equation can reduce to a simple form at present ($a=1$),
\beq
 h_W \simeq h \left[ 1 - \frac{5}{6} \Omega_{\rm m} \frac{\delta_{{\rm b}0}}{D_0} \right].
\eeq
The other cosmological parameters have similar relations as,
\begin{align}
 \Omega_{{\rm m}W} &\simeq \Omega_{\rm m} \left[ 1 + \frac{5}{3} \Omega_{\rm m} \frac{\delta_{{\rm b}0}}{D_0} \right], \notag \\
 \Omega_{\Lambda W} &\simeq \Omega_\Lambda \left[ 1 + \frac{5}{3} \Omega_{\rm m} \frac{\delta_{{\rm b}0}}{D_0} \right], \notag \\
 \Omega_{{\rm K}W} &= 1- \Omega_{{\rm m}W} - \Omega_{\Lambda W} \simeq 1 - \left( \Omega_{\rm m} + \Omega_\Lambda \right) \left[ 1 + \frac{5}{3} \Omega_{\rm m} \frac{\delta_{{\rm b}0}}{D_0} \right].
\end{align}
In our case ($\delta_{{\rm b}0} = \pm 0.01$), these SU parameters are given in Table \ref{table1}. 
}


\section{Relation between 3D and 2D power spectra}
{
We consider the 3D and 2D Fourier transforms of density field in a cubic box of side length $L$. The Fourier component and power spectrum for 3D density filed $\delta(x,y,z)$ are given as
\begin{align}
 \widetilde{\delta}(k_x,k_y,k_z) &=\int_0^L \!\! dx \int_0^L \!\! dy \int_0^L \!\! dz \, \delta(x,y,z) \, {\rm e}^{i(k_x x+k_y y+k_z z)}, \notag \\
 P(k) &= \frac{1}{V} \langle | \widetilde{\delta}(k_x,k_y,k_z) |^2 \rangle,
\label{eq:ap1}
\end{align}
where $V=L^3$ is the volume, $k=(k_x^2+k_y^2+k_z^2)^{1/2}$, and the wavevector components have discrete values, i.e., $k_{x,y,z}=(2\pi/L) \, n_{x,y,z}$ with integers $n_{x,y,z}=0, \pm 1, \pm 2, \cdots$.
Next, the density filed projected onto $xy$ plane is given as
\beq
 \delta^{\rm 2D}(x,y) = \frac{1}{L} \int_0^L \!\! dz \, \delta(x,y,z).
\label{eq:ap2}
\eeq
Then, the 2D Fourier component and power spectrum are
\begin{align}
 \widetilde{\delta}^{\rm 2D}(k_x,k_y) &=\int_0^L \!\! dx \int_0^L \!\! dy \, \delta^{\rm 2D}(x,y) \, {\rm e}^{i(k_x x+k_y y)}, \notag \\
 P^{\rm 2D}(k) &= \frac{1}{S} \langle | \widetilde{\delta}^{\rm 2D}(k_x,k_y) |^2 \rangle,
\label{eq:ap3}
\end{align}
where $S=L^2$ is the surface area and $k=(k_x^2+k_y^2)^{1/2}$.
Therefore, from equations (\ref{eq:ap1})-(\ref{eq:ap3}), we have a relation between 3D and 2D quantities as
\begin{align}
 \widetilde{\delta}^{\rm 2D}(k_x,k_y) &= \frac{1}{L} \widetilde{\delta}(k_x,k_y,k_z=0), \notag \\
 P^{\rm 2D}(k) &= \frac{1}{L} P(k).
\end{align}
}

\bsp	
\label{lastpage}
\end{document}